\documentclass[twocolumn]{aastex631}
\usepackage{hyperref}
\usepackage[normalem]{ulem} 

\usepackage{graphicx} 
\usepackage{natbib} 
\usepackage{amsmath}
\usepackage{amssymb} 
\usepackage{fancyhdr} 
\usepackage{latexsym}
\usepackage{color} 
\usepackage{txfonts}
\definecolor{mypurple}{RGB}{128,0,255}

\graphicspath{{./}}
\usepackage{soul}
\usepackage{aas_macros}
\usepackage{epstopdf}
\usepackage{mydefs_iris2}
\usepackage[normalem]{ulem}

\def\asd#1{\textcolor{black}{#1}}

\begin{document}

\title{The IRIS$^{2+}$ inversion tool: recovering the radiative losses and the thermodynamics in the lower solar atmosphere}

\author[0000-0002-3234-3070]{Alberto Sainz Dalda}
\affil{Lockheed Martin Solar \& Astrophysics Laboratory, 3251 Hanover Street, Palo Alto, CA 94304, USA}
\affil{Search for Extraterrestrial Intelligence (SETI) Institute, 339 Bernardo Ave, Suite 200, Mountain View, CA 94043, USA}
\author[0000-0002-4640-5658]{Jaime de la Cruz Rodr{\'i}guez}
\affil{Institute for Solar Physics, Dept. of Astronomy, Stockholm University, AlbaNova University Centre, SE-106 91 Stockholm, Sweden}
\author[0000-0003-0975-6659]{Viggo Hansteen}
\affil{Lockheed Martin Solar \& Astrophysics Laboratory, 3251 Hanover Street, Palo Alto, CA 94304, USA}
\affil{Search for Extraterrestrial Intelligence (SETI) Institute, 339 Bernardo Ave, Suite 200, Mountain View, CA 94043, USA}
\author[0000-0002-8370-952X]{Bart De Pontieu}
\affil{Lockheed Martin Solar \& Astrophysics Laboratory, 3251 Hanover Street, Palo Alto, CA 94304, USA}
\affil{Rosseland Center for Solar Physics, University of Oslo, P.O. Box 1029 Blindern, NO-0315 Oslo, Norway}
\affil{Institute of Theoretical Astrophysics, University of Oslo, P.O. Box 1029 Blindern, NO-0315 Oslo, Norway}
\author[0000-0002-5879-4371]{Milan Go\v{s}i\'{c}}
\affil{Lockheed Martin Solar \& Astrophysics Laboratory, 3251 Hanover Street, Palo Alto, CA 94304, USA}
\affil{Search for Extraterrestrial Intelligence (SETI) Institute, 339 Bernardo Ave, Suite 200, Mountain View, CA 94043, USA}



\begin{abstract}

 We introduce an improved and fast inversion tool that is able to provide the thermodynamics of the solar atmosphere from the photosphere to the top of the chromosphere, as well as the integrated radiative losses in the chromosphere for data observed by the Interface Region Imaging Spectrograph (IRIS). This NASA mission has been observing the Sun and providing, among other kinds of data, multi-line spectral observations sensitive to changes in the lower solar atmosphere since 2013. In this paper, we explain the new inversion tool \irissqp\ based on the \irissqp\ database, which is based on 135,472 synthetic representative profiles (RP), each of them consisting of 6 chromospheric lines and 6 photospheric lines, their corresponding representative model atmospheres (RMA), and the integrated radiative losses (IRL) associated with these atmospheres. A nearest neighbor (k-nn) model algorithm is trained with the synthetic representative profiles to {\it predict} the closest RP in the database to the one observed, at which point \irissqp\ assigns the RMA and the IRL to the location of that observed profile.  We have compared the results obtained by \irissqp\ with results obtained from the state-of-the-art inversion code STiC, which is also used to build the \irissqp\ database. We find that the thermodynamics and the IRL obtained with both methods are comparable in most cases. Therefore, \irissqp\ is a fast and reliable inversion tool that provides approximate values of the thermodynamic state and the radiative losses in the lower solar atmosphere for a large variety of solar scenes observed with IRIS.
 
\end{abstract}



\section{Introduction} \label{sec:intro}

Inferring the physical state of the chromospheric plasma from observations remains an outstanding challenge for solar physicists. Due to low collisional rates in the chromosphere, a non-local thermodynamical equilibrium description of the radiation field is required to explain the emerging intensities that we observe (see, e.g., \citealt{Vernazza81}). As a result, there is no straightforward translation from observed intensities to chromospheric physical parameters, as it might be the case in the photosphere. Inversion methods (inversions hereafter) attempt to model the transport of radiation through a model atmosphere in order to fit the observations \citep[see review by][]{delToroIniesta16}. Inversions are one of the most powerful quantitative tools for chromospheric data as they maximize the number of physical parameters that can be recovered if sufficiently sensitive data are included: temperature, line-of-sight velocity, microturbulence, and the magnetic field vector. 

Results from modern non-LTE inversion codes \citep{Socas-Navarro15,2018A&A...617A..24M,delaCruzRodriguez19,RuizCobo22} have been presented for almost 20 years now \citep[e.g.,][]{Pietarila07c,delaCruzRodriguez13a,2018ApJ...860...10K,2025A&A...696A.105S}. But more recently, the fidelity of results from inversions has significantly increased thanks to the inclusion of more spectral lines that increase the depth resolution \citep[e.g.,][]{2019ApJ...870...88E}. The latter is allowing us to perform more complex post-processing analysis, such as the calculation of the divergence of the radiative flux to estimate the net radiative losses in the chromosphere \citep{DiazBaso21, Morosin22, Yadav22,2024ApJ...976...21D}. Knowing the net radiative losses in the chromosphere is important because they pose a lower limit of the total amount of energy that must be replenished by heating mechanisms at any given time in the chromosphere. 

Most implementations of inversion techniques attempt to reconstruct the observed spectral lines at a given location with a single model atmosphere, assuming that observations in all spectral windows are co-spatial, co-temporal, and acquired with a similar spatial resolution. Although there is some tolerance for misalignment in space and time, they can cause artifacts in the reconstructed parameters \citep{2018A&A...614A..73F,2019A&A...627A.101V}. The IRIS mission obtains multi-line datasets using a slit-spectrographs, naturally fulfilling the spatio-temporal alignment criteria.

In this study, we use the multi-line and multi-layer inverted data of observations taken with the Interface Region Imaging Spectrograph \citep[IRIS,][]{DePontieu14a} and present a
fully functional inversion tool that provides the thermodynamics, i.e., the profile (as a function of optical depth) of temperature ($T$),  line-of-sight velocity ($v_{los}$), turbulent motions ($v_{turb}$, also called micro-turbulence velocity, $v_{mic}$), and electron density ($n_e$), as well as the integrated radiative losses ($IRL$). To this aim, we use an extended version of the $IRIS^{2+}$ database developed by \cite{SainzDalda24}. In Section \ref{sec:data}, we summarize how this database was built and extended for this investigation. In Section \ref{sec:inversion} we describe the \irissqp\ inversion tool. \irissqp\ provides, in addition to the thermodynamics, the integrated radiative loss (IRL) for a given IRIS data set. We explain how we have computed the IRL for the representative model atmospheres ($RMA$) of the \irissqp\ database in Section~\ref{sec:irl}. A statistical comparison between the results obtained by \irissqp\ and the STockholm inversion Code \citep[STiC][]{delaCruzRodriguez16, delaCruzRodriguez19} is discussed in detail in Section \ref{sec:comparison}. 
The results are discussed in Section \ref{sec:discussion}, while the conclusions of this investigation are given in Section \ref{sec:conclusions}. In Appendix \ref{appendix:comparison_IRL}, we compare the integrated radiative losses calculated whit all the species considered together in non-LTE with the integrated radiative losses calculated with the species individually, i.e., one species in non-LTE and the others in LTE. In Appendix \ref{appendix:detailed_comparison_chi2}, we show a detailed study of the quality of the fit of the inversion ($\chi^{2}$) when different combinations of weights are given to the spectral lines in the inversions.

\section{Data}\label{sec:data}

The \irissqp\ database contains the following spectral lines: the chromospheric lines \cii, the \mgii, and the \mguv; and the photospheric lines Ti \iifns\ 2785.46 \AA, Fe \ifns\ 2793.22 \AA, Fe \ifns\ 2809.15 \AA, C \ifns\ 2810.58, Ni \iifns\ 2815.18 \AA, and Fe \ifns\ 2827.33 \AA.\footnote{For the sake of simplicity, in this paper, we will refer to this set lines as the {\it \irissqp\ lines}.} 
The first version of the \irissqp\ database presented by \cite{SainzDalda24} was created with 40,320 \rps\ and their corresponding \rmas. In order to address the added complexity of considering the entire lower solar atmosphere from photosphere to transition region, this database has been increased to roughly 135,000 \rp-\rma\ pairs. The datasets selected to extend the \irissqp\ are treated in the same way as in the previous version: A dataset is pre-sorted before obtaining the \rps\ into one of five main classes: umbra, pore-like, penumbra, quiet sun, or plage. Thus, a fixed number of \rps\ is guaranteed for each of these classes. The \rps\ are calculated using the unsupervised clustering {\it k-means} method \citep{Steinhaus57, MacQueen67}. Once the multi-line \rps\ are obtained, they are inverted using the STiC code, as explained in detail in Section 2.3 of \cite{SainzDalda24}. 

This new version of the \irissqp\ database includes data from active regions as well as other sources. However, as in the case of \irissq, it is important to note that \irissqp\ does not consider highly energetic events such as flares and surges, and does not include all of the sometimes unusual spectral profiles found on the Sun (e.g., Ellerman bombs). The main reason these kinds of data are avoided is the difficulty in inverting their profiles, which often show very broad and/or pointy shapes \citep[see][]{SainzDalda23}. Profiles from episodic high-energy events require a devoted inversion scheme that allows one to fit the profiles while avoiding too many free parameters and providing realistic and interpretable thermodynamic models. Special attention must be paid to the large variation of the relative line intensities during such energetic events \citep{SainzDalda23,Roy24}. That strong variation seen in the lines studied by these authors, the \cii\ lines and the \mgii\ lines, reflects the dynamic response these events cause through the chromosphere. 

The lines chosen in this study have been selected because they are sensitive to changes in the thermodynamic conditions in different regions of the lower solar atmosphere, more precisely, from the high chromosphere to the mid photosphere. In total, for 550 data sets, 320 multi-line \rps\ were calculated per data set, and then inverted using STiC. Figure \ref{fig:distribution_data} shows the location on the solar disk (left) 
of these data sets and the distribution as a function of the years of the IRIS mission (right).

\begin{figure*}
    \centering
    \includegraphics[width=1\linewidth]
    {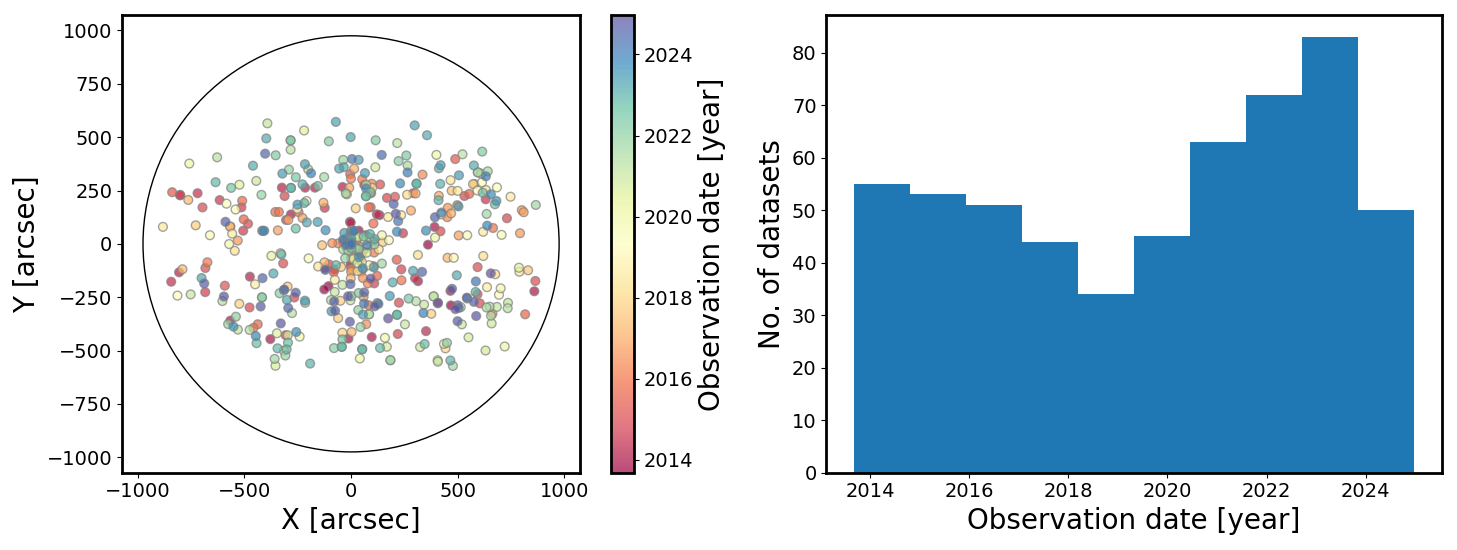}
    \caption{Left: location of the 550 datasets selected to build the \irissqp\ database. Right: distribution of the observation date of these datasets during the years of the IRIS mission (which was launched in 2013).}
    \label{fig:distribution_data}
\end{figure*}

Once all the \rps\ are inverted, those profiles with a {\it very bad fit} are discarded. To determine the quality of  a fit, we use the $\chi^2$ metric, defined as:

\begin{equation}\label{eq:chi2}
 \chi^{2} = \frac{1}{\nu}\sum_{i=1}^{q}{(I(\lambda_i)^{obs} - I(\lambda_i, \mathbf{M})^{syn})^{2}\frac{w_{i}^2}{\sigma_{i}^2}}, 
\end{equation}

\noindent
with $i = 1,..., q$ the sampled wavelengths, $w_{i}$ their weights, $\sigma_{i}$ uncertainties of the observation,
 and $\nu$ the number of degrees of freedom, that is, the difference between the number of observables (q) and the number of free parameters in the model $\mathbf{M}$ or of the nodes during the inversion. Because of computational efficiency, STiC uses a unique value of $w_{i}/\sigma_{i}$ for each data set. Different weights are given for different spectral ranges, and we assume that $\sigma_i$ is the same for all $i$ in each spectral channel. That is, we have used $w_{\textrm{C {\footnotesize II}}}/\sigma_{\textrm{C {\footnotesize II}}}$
 , $w_{\textrm{\mguv}}/\sigma_{\textrm{NUV}}$, $w_{\textrm{\mgii}}/\sigma_{\textrm{NUV}}$, and $w_{\textrm{photos}}/\sigma_{\textrm{NUV}}$.

 Due to computational constraints, only one vector of weights is used for all the \rps\ of a dataset. The weights of the datasets are obtained by computing the most frequent value of the intensity in a spectroheliogram map in the Mg II h and k line cores, and setting the weight of the \cii\ lines by adjusting their intensity to that of the \mgii\ lines. In this way, they contribute evenly to the $\chi^2$. For the other lines, we apply the relative weights $w_{\textrm{\mguv}}$:$w_{\textrm{\mgii}}$:$w_{\textrm{photos}}$ = 3:1:2. A more detailed discussion about the important role played by the weights is discussed in Appendix \ref{appendix:detailed_comparison_chi2}.

To summarize, the new extended version of \irissqp\ consists of 135,472 \rps, sampled in 944 wavelength bins, their corresponding \rmas, i.e., the temperature T, line of sight velocity \vlos, turbulent velocity \vturb, and electron density \nne\ evaluated at 39 optical depths from $-7.6\leq~$\ltau$~
\leq0$, and the representative integrated radiative loss (\rirl) for each \rma. 

\section{Inversion method}\label{sec:inversion}

The \irissqp\ inversion tool follows the same approach as the IRIS Inversion based on Representative Profiles Inverted by STiC \citep[\irissq,][]{SainzDalda19}. For a given observed profile, these tools look for the closest synthetic profile in a database and assign the model atmosphere that produces that synthetic profile to this profile. The database is, in both cases, built from the inversion of the {\it representative profile} ($RP$) with the STiC code. An RP is the average (centroid) of those spectral profiles that share a similar shape, and therefore that have been generated under similar atmospheric conditions. The thermodynamic values recovered from the inversion of an $RP$ are representative of the profiles associated with that $RP$, and so is their {\it representative model atmosphere} ($RMA$). Both \irissq\ and \irissqp\ assign the closest synthetic $RP$ and its corresponding \rma\ to the location of the observed profiles. In addition, \irissqp\ also assigns the \rirl\ to that location. 

In \irissq, the closest synthetic profile is calculated by using the Euclidean weighted distance. 
In \irissqp\ we use the same metric, but we take advantage of the {\it k-nearest neighbor} algorithm \citep[k-nn,][]{Fix51Jr, Cover67}. This algorithm allows for predicting the $k$ closest samples in a training dataset for an input sample by training the similarity of the samples in that model. In our case, we use the k-nn algorithm to train the similarity between the \rps\ of the \irissqp\ database, and for a given external sample, find the $k$ closest (most similar) elements in the database. In \irissqp, we use $k=1$, therefore it finds the closest $RP$ in the \irissqp\ database. Once the model is trained, the search for the closest synthetic \rps\ to the observed profiles is very fast.

While in \irissq\ the weights are important, they are mostly used to block (weight with zero) some spectral regions. This is because the {\it important} lines in the \irissq\ database are the \mgii\ lines and the second and third lines of the \mguv\ (\mguvtt). The intensity of the \mgii\ lines is similar and no relative weights are needed, while the \ion{Mg}{2} UV-triplet lines usually need a relative weight with respect to the \mgiik\ lines to compensate for the difference between their intensities. 

In the case of \irissqp, because we can use any combination of the lines in the database and they have very different relative intensities, the values given to the weights during the inversion are 
a critical component in fitting the observations. \irissqp\ offers several options to deal with this situation. 

The ``{\tt mean}'' option, for a given line, \irissqp\ calculates the mean value of the intensity map\footnote{\irissqp\ automatically calibrates the data to the specific intensity units, i.e., $\mu$erg~cm$^{-2}$~s$^{-1}$~sr$^{-1}$~\AA$^{-1}$.}, obtained as the integrated value in a spectral range around the core for that line. Then, the weights are determined as the ratio between these mean values with respect to the mean value of the integrated intensity map at \mgiik. 

The ``{\tt default}'' option provides the following combination of weights: $w_{\textrm{\cii}}$:$w_{\textrm{\mguv}}$:$w_{\textrm{\mgii}}$:$w_{\textrm{photos}}$ = 100:3:1:2. 

In the ``{\tt array}'' option, the weights are passed as an array of weight values for each spectral line. If the values are floating-point numbers, the weights are passed as given in the array. If the numbers are integers, those values with a value equal to 0 are weighted with 0.0 (discarded). For those values in the array with an integer different than 0, the weights passed are the ones corresponding to the {\tt 'mean'} case. Finally, there is a ``{\tt manual}'' option that allows the user to pass weights through an interactive window. For all options, the weights are taken into account before training the k-nn model. 

It is worth noting that the \rma\ provided by \irissqp\ contains information along the full optical depth that is used during inversion using the 12 lines that are used for \irissqp. Thus, the user should consider with a critical eye values of the thermodynamics for those optical depths where the 12 lines are not sensitive to local thermodynamics, or where the lines that were weighted with 0 - or were not available in the original IRIS data set - are sensitive to changes in the thermodynamics. 
For example, if the user sets the weights to 0 for photospheric lines (or those lines were not present in the data), then the thermodynamic values between -2 $\lessapprox$\ltau$\lessapprox$ 0 resulting from the inversion should not be considered reliable since they are poorly or not constrained by the observations.  

Due to the spectral versatility of the IRIS observations, in which the variety of data containing some or all of the lines considered by \irissqp, and differing observational setups, including especially the spectral sampling, as well as the varying nature of the dynamic events observed, the inversion code needs to train the k-nn model with the database and the observed data in such a manner that they share the same (observed) lines, the same spectral sampling, and the same weights. Therefore, the k-nn model is trained every time for the observation to be inverted. Thus, the \rps\ of the \irissqp\ are {\it transformed} to the observation's spectral sampling and {\it cropped} into the spectral ranges used in the observed data. Then, the new database is used to train the k-nn model, and with this model, the closest transformed \rp\ to the observed one is found.

\begin{figure*}
    \centering
    \includegraphics[width=1\linewidth]
    {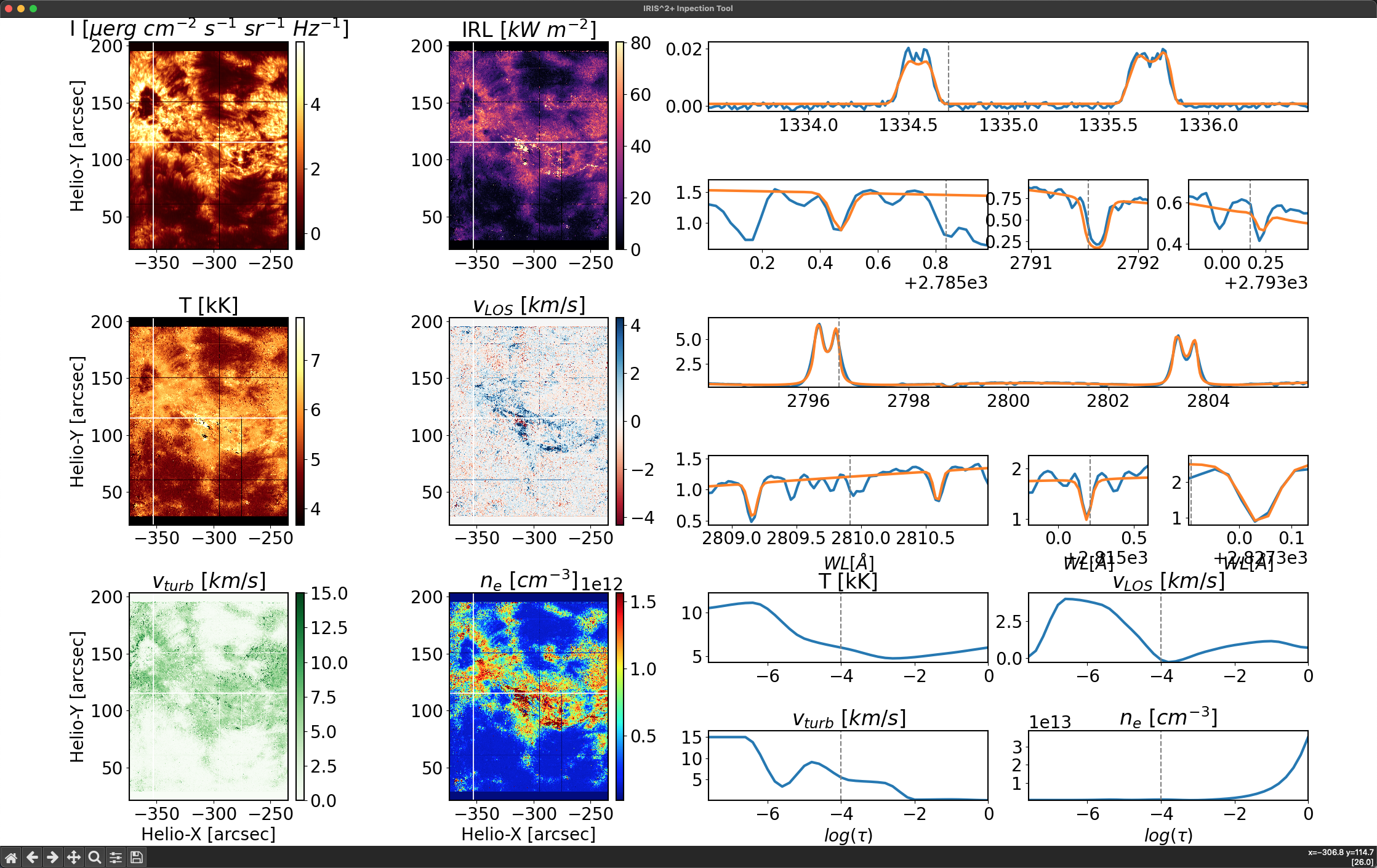}
    \caption{Screenshot of the interactive \irissqp\ inspection tool. In the first and second columns the map images show from top to bottom, left to right: the specific intensity in \radunits at the wavelength ($\AA$) selected in one of the spectral plots shown in the third column (first 8 plot panels from the top); the integrated radiative loss (IRL) in $kW~m^{-2}$; the T in $kK$; the \vlos\ in $km~s^{-1}$; the \vturb\ in $km~s^{-1}$; and the \nne\ in $cm^{-3}$. These thermodynamic variables show the value at the optical depth (\ltau) determined by the positions in the thermodynamic plots displayed in the last four panels in the third column. The spectral plots show the observed spectral profile (blue) and the synthetic spectral profile found by \irissqp\ (orange) that best fits it. The specific intensity shown in the first top panel can be toggled to show either the observed intensity, as it is in the current figure, or the synthetic intensity corresponding to the synthetic best fit profiles at that particular wavelength. Similarly, the IRL panel can be toggled to show the $\chi^2$ map. In all cases, the value below the cursor these cases is shown in the bottom right corner of the window.}
    \label{fig:iris2p_vistool}
\end{figure*}

The results provided by \irissqp\ can be easily visualized and inspected in the {\it \irissqp\ Inspection Tool}. Figure \ref{fig:iris2p_vistool} shows this tool for the inversion of the data set taken by IRIS on August 15, 2014, at 22h:36min:09s (UT).

\section{The Integrated Radiative Loss}\label{sec:irl}

A detailed solution for the radiative loss requires the computation of the divergence of the radiative flux $F$ \citep{Uitenbroek02,Rutten03} as:
\begin{equation}\label{eq:qloss}
    Q = \nabla F = \frac{1}{2}\int_{-1}^{1}\int_{0}^{\inf} \alpha_{\nu}(z,\mu)[S_{\nu}(z,\mu)-I_{\nu}(z,\mu)]d\nu d\mu
\end{equation}

\noindent
 with $\alpha_{\nu}$ being the absorption coefficient, $S_{\nu}$ the source
function, and $J_{\nu}$ the mean intensity over solid angle. 

The latest version of STiC allows the computation of the radiative losses from an atmospheric model for a set of radiating species. STiC uses the RH code \citep{Uitenbroek01} to make these calculations. STiC has been recently used to calculate the radiative loss in different solar features, such as a reconnection event due to a magnetic flux cancellation observed in the photosphere \citep{DiazBaso21}, a plage region \citep{Morosin22}, a C-class flare \citep{Yadav22}, or the quiet-Sun \citep{2024ApJ...976...21D}.

Here, we present a database of {\it Representative Integrated Radiative Loss} (RIRL) obtained from the \rmas\ in \irissqp. The radiative losses associated with an \rma\ were calculated considering the largest contributors to the radiative loss in the chromosphere. Following \cite{Vernazza81}, these contributors are:
\begin{itemize}
    \item Contributors from H: Lyman-$\alpha$, H-$\alpha$, H-$\beta$, the Lyman continuum, and the Balmer continuum.
    \item Contributors from \ion{Mg}{2}: \mgii\ lines, and the \mguv.
    \item Contributors from \ion{Ca}{2}: \ion{Ca}{2} H \& K lines, and the \ion{Ca}{2} IR triplet.
\end{itemize}

We ignore the contribution to the radiative loss from the photosphere, which is mostly due to the $H^{-}$ continuum and the many weak photospheric lines, \asd{including the \ion{Fe}{2} lines. The contribution of \ion{Fe}{2} lines to the radiative losses can be significant, particularly in the lower chromosphere and photosphere, where they may account for up to 50\% of the total radiative losses \citep{Anderson89}. However, these authors considered a simplified atomic model where many levels were grouped, forming meta-levels with larger oscillator strength, which leads to transitions with an unrealistically large opacity (due to the grouping of levels) that increases the interaction with the radiation field. 
Consequently, the impact of Fe II lines on the IRL must be revisited using a more realistic and comprehensive atomic model for Fe {\footnotesize I, II, III}. Constructing such a model —given the very large number of transitions and the complexity involved— is beyond the scope of the present work.} 

\begin{figure*}
    \centering
    \includegraphics[width=0.95\linewidth]
    {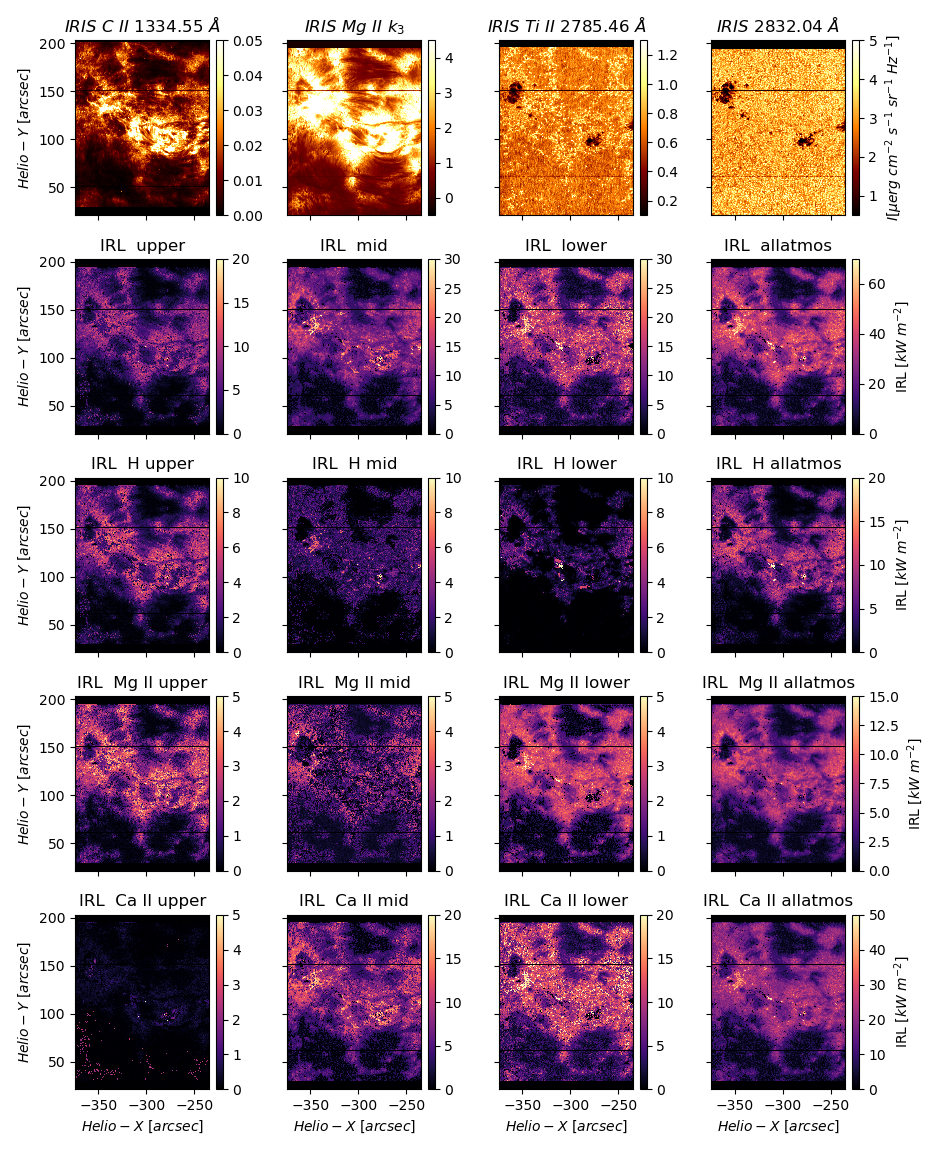}
    \caption{First row: Integrated slit-reconstructed intensity map around \ciifour, \mgiik$_3$, \ion{Ti}{2} 2785.46 $\AA$, and 2832.04 $\AA$, roughly corresponding to the high chromosphere, mid chromosphere, the high photosphere, and the continuum photosphere, respectively. Second to the fith row: Total contribution (second row) and partial contribution from the main ions (H, Mg II, and Ca II in the third, fourth, and fifth rows respectively) to the integrated radiative losses in three regions in the lower solar atmosphere (first, second, and third columns respectively) and the total integrated radiative loss (fourth column).}
    \label{fig:irl}
\end{figure*}

We have used the latest version of STiC to calculate the radiative losses. This code, using the RH code in the background, now has the capability to evaluate and return the losses described by Equation \ref{eq:qloss}.
The total integrated radiative loss is then calculated by integrating along the optical depth through the chromosphere. We define the chromosphere as the region defined by the optical depth range where the $4\leq T~[kK]\leq18$. This range was determined from the mean and the standard deviation of the $T(\tau)$ in the \irissqp\ database. Although a more restrictive range ($4.5\leq T~[kK]\leq12$) provides almost similar results to the ones obtained in the comparison studied in this paper, we have used the former one to account for those events that may release or absorb more energy.

Ideally, the calculation of the IRL from a given \rma\ should consider the contributions of all the contributors simultaneously. However, when we obtained the \rma\ for the \irissqp\ database, we could not invert the observed \rps\ taking into account simultaneously the hydrogen, \ion{C}{2}, and the \ion{Mg}{2} transitions in non-LTE. Thus, during the inversion, we had to consider only the \ion{C}{2} and the \ion{Mg}{2} transitions simultaneously in non-LTE, since including hydrogen made STiC convergence difficult or, in many cases, impossible because of a code crash. As a consequence, the inversion was either extremely computationally expensive or not possible. We therefore decided to compute the IRL considering these species independently, which means that for each set of transitions, only that one element is active (under non-LTE) while the other two are passive (computed in LTE). RH includes the contribution of bound-free (B-F) transitions from active and non-active (LTE) atoms. Several B-F transitions from the three atoms considered in this study overlap in different parts of the spectrum, affecting the overall opacity and emissivity of the continuum. For example, considering Hydrogen in LTE when converging the \ion{Mg}{2} can influence the UV background opacity and affect the final result for the calculation in the \ion{Mg}{2} atom. Given that we used an LTE equation-of-state, and that the hydrogen B-F opacity was also calculated in LTE during the inversion, we do not think that this way of proceeding is a dominant source of error in our calculation of the radiative losses. In Appendix \ref{appendix:comparison_IRL}, we show the comparison between calculating the IRL considering all the species together and independently for a data set included in the \irissqp\ database. There, we have quantified that the radiative losses computed considering the species in non-LTE independently overestimate by $\approx$15\% the radiative losses computed with all the species in non-LTE together. This overestimation corresponds to the radiative losses integrated over the whole chromosphere. However, they show a very similar behavior in the mid chromosphere.

Following the same approach as \cite{Morosin22}, Figure \ref{fig:irl} shows the partial contribution of these lines to the integrated radiative losses in three equal-sized optical depth ranges in the chromosphere. In other words, we divide into three parts the optical depth range that is obtained by imposing the temperature range used to define the chromosphere. This range in optical depth is RMA-dependent, and so are the three regions. We refer to these regions as {\it upper, mid}, and {\it lower} in  Figure \ref{fig:irl}. The figure helps us to assess the contribution in height to the chromospheric radiative losses from each of the main contributors. Although we have computed these values for all the \rmas\ in the \irissqp database, and they are included in the \irissqp\ inversion tool, to simplify the comparison, in this study we use the total integrated radiative losses ($IRL$) shown in Figure \ref{fig:irl}, in the panel titled {\it ``IRL allatmos''},  with $IRL = IRL^{upper}+IRL^{mid}+IRL^{lower}$. \irissqp\ provides the result obtained by evaluating Equation \ref{eq:qloss} using STiC.

\section{Comparison between the ST\lowercase{i}C and \irissqp\ inversion results}\label{sec:comparison}

\begin{table*}[!htp]
    \centering
    \begin{tabular}{ccccccc}
        Id. No. & IRIS OBSID &  Longitude & Latitude & $\mu$ & Exp. time & X scale  \\
         & &  [$''$] & [$''$] &  &  [$s$] & [$''/px$]  \\
         \hline
        \#0 & 20140815\_223609\_3880012196 &  -304.39 & 112.14 & 0.94 & 30 &  0.35 \\
        \#1 & 20160301\_232931\_3680012175 &  -71.68 & -328.30 & 0.94 & 30 &  0.35 \\
        \#2 & 20140404\_051620\_3880012197 &  117.38 & 367.75 & 0.92 & 30 &  0.35 \\
        \#3 & 20140211\_212831\_3880012191 &  291.91 & 213.20 & 0.93 & 30 &  0.35 \\
        \#4 & 20240910\_044829\_3893010094 &  -207.19 & 138.81 & 0.97 & 15 & 2.00 \\
        \#5 & 20241003\_062122\_3893010094 &   -5.54 & -365.22 & 0.92 & 15 & 2.00 \\
        \#6 & 20160810\_065848\_3882010194 &  129.15 &  39.79 & 0.99 & 15 &  2.00 \\
        \#7 & 20240310\_040941\_3893010094 &  -242.55 & 247.75 & 0.93 & 15 & 2.00 \\
        \#8 & 20220628\_045806\_3893010094 &  122.60 & -277.58 & 0.95 & 15 & 2.00 \\
        \#9 & 20221113\_051649\_3893010094 &  410.02 & 204.62 & 0.88 & 15 &  2.00 \\
    \end{tabular}
    \caption{Main observational features of the IRIS data sets selected for the comparison study. The spectral scale is $13~m\AA/px$ for the \cii\ lines and $25~m\AA/px$ for the rest of the lines in all the data sets. The scale in the Y axis (along the slit) is $0.17~''/px$ in all the data sets. $\mu$ is the cosine of the heliocentric viewing angle $\theta$.}
    \label{tab:infobs}
\end{table*}

To quantify the quality of \irissqp, we have inverted 10 IRIS data sets both with STiC and with \irissqp. These observations include all the lines represented in the \irissqp\ database. Figures \ref{fig:intmap_1} and \ref{fig:intmap_2} show the selected data in the high chromosphere (first column), mid chromosphere (second column), and upper photosphere (third column). Our selection is based on the variety of solar features observed in the chromosphere. 
Table \ref{tab:infobs} shows the IRIS observation identifier (IRIS OBSID) and some of the observing characteristics for the selected data sets.

%

Due to the large computational time needed to invert all the profiles in the selected observations, for the masked area of each dataset, we have only inverted 1 px every other 16 px, i.e., the left, bottom pixel of a $4\times4$~px$^{2}$ square. The masked area is the common area in the far- and near-ultraviolet (FUV and NUV, respectively) IRIS detectors covered by solar radiation, after being co-aligned. Thus, for example, a dataset that originally recorded $1092\times64$~px$^{2}$ (69,888~px), the common, co-aligned mask covered $\approx~67,000$~px, and we inverted $\approx~4,200$~px. The inversion of the datasets using \irissqp\ is for all the pixels in the common, co-aligned masked area. Subsequently, the locations of the profiles inverted by STiC are selected for the comparison.

\begin{figure*}
    \centering
    \includegraphics[width=1\linewidth]
    {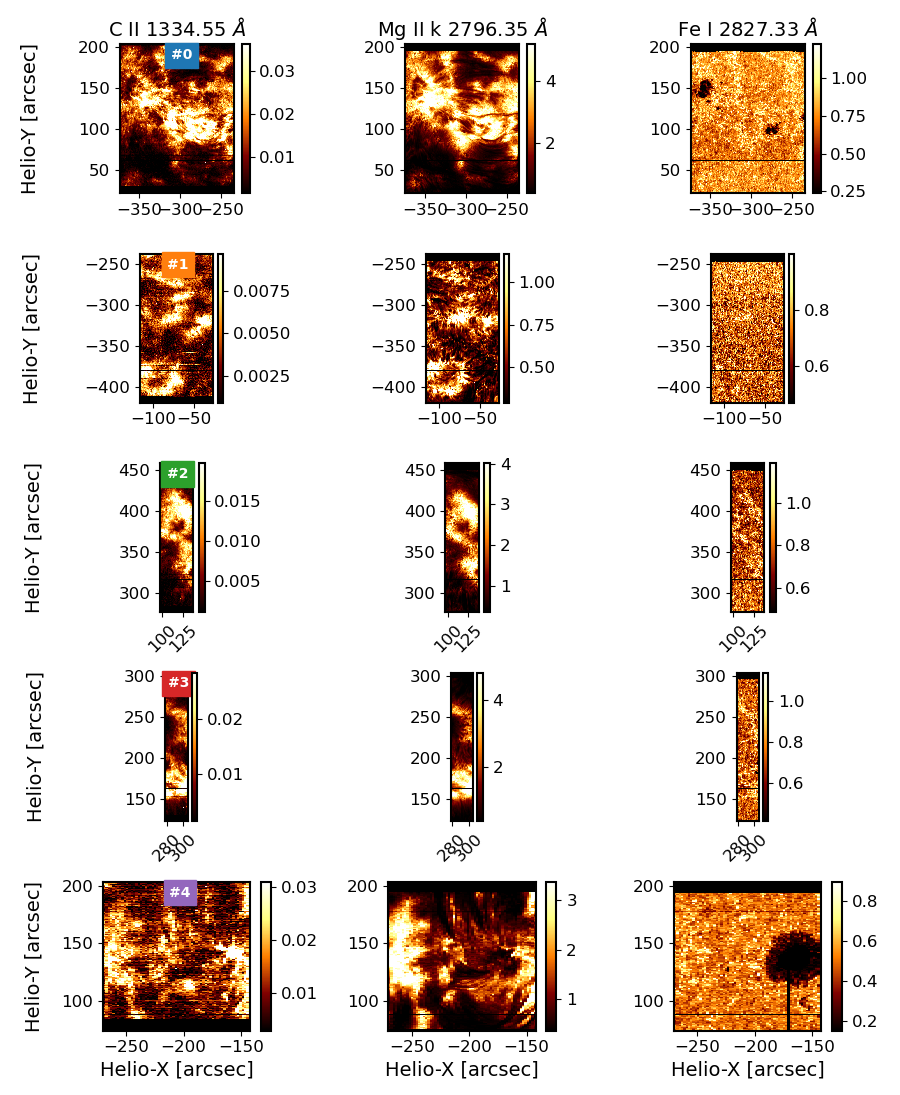}
    \caption{The first 5 IRIS observations used for the comparison between STiC and \irissqp. The first column shows the intensity in the photosphere, the mid chromosphere, and high chromosphere as observed in the core of Fe \ifns\ 2827.33, \mgiik, and \ciifour\ respectively. The images show the specific intensity in \radunits in these wavelengths.}
    \label{fig:intmap_1}
\end{figure*}
\begin{figure*}
    \centering
    \includegraphics[width=1\linewidth]
    {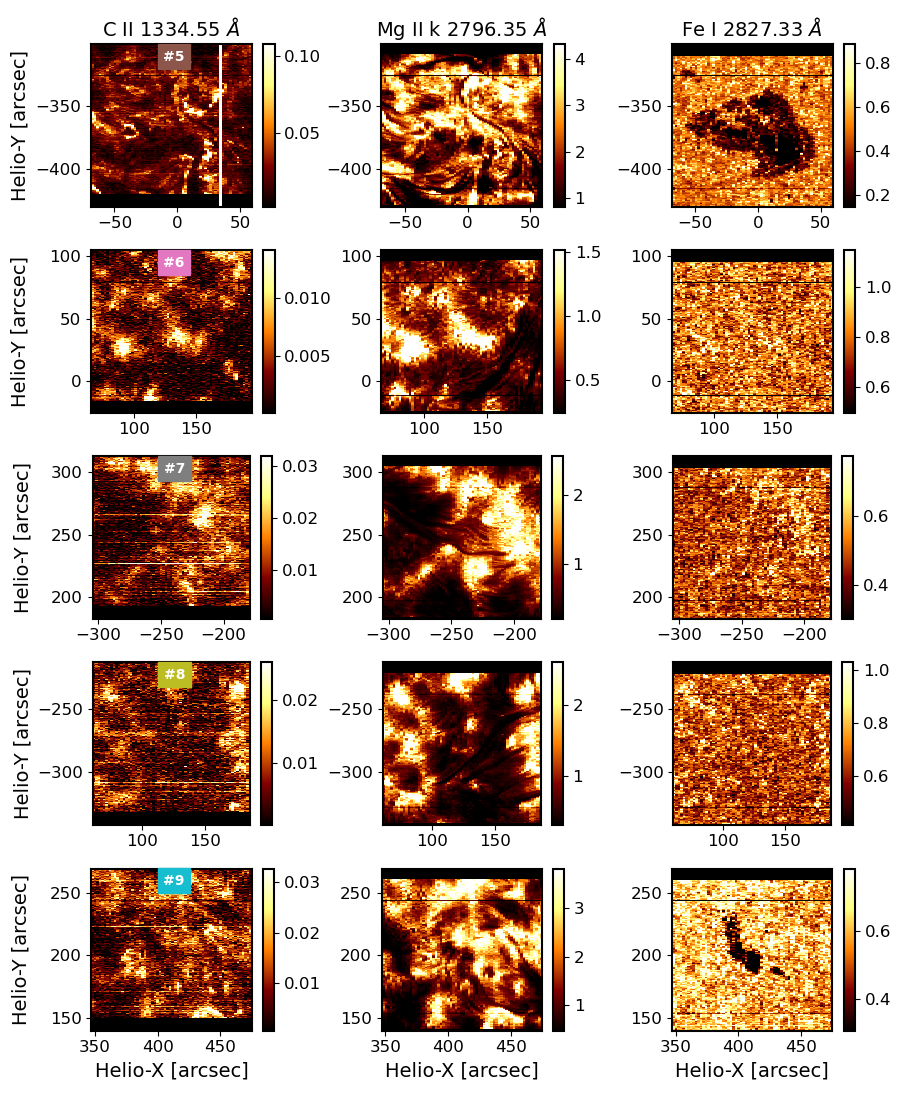}
    \caption{Same as figure \ref{fig:intmap_1} for the last 5 IRIS observations used for the comparison between STiC and \irissqp.}
    \label{fig:intmap_2}
\end{figure*}

One might consider the STiC solution as the {\it ground truth}; however, as \citet{SainzDalda19} pointed out, for a given observed profile, the closest profile found in the database may fit better than the solution obtained from a dedicated inversion of that observed profile. This is because the dedicated inversion tries 5 independent inversions from 5 different initializations and returns the best solution. 
This is the inversion scheme used to build the database from the RPs. Nevertheless, \irissqp\ database may have more than one synthetic profile very similar to the observed one, each one having been obtained after trying 5 different initialization paths, therefore increasing the pool of possible close fits. 

We have focused our comparison on the linear correlation between the IRL and the thermodynamic variables obtained from the STiC and the \irissqp\ inversions, when all the lines of the \irissqp\ are considered and weighted with the same weights used by STiC. However, thanks to the variety of spectral lines recorded in the IRIS observations, and the versatility of \irissqp, the user may want to fit only a selection of lines, change the weight of some lines with respect to others, or even ignore some of them. For this reason, Appendix \ref{appendix:detailed_comparison_chi2} contains a detailed study to quantify the impact on the inversion results due to the potential user's lines selection.
 
\subsection{Comparison of the integrated radiative loss}\label{sec:comparison_irl}

Figure \ref{fig:lincorr_irl} shows the comparison between the integrated radiative loss obtained by STiC (IRL$_{STiC}$) and by \irissqp\ (IRL$_{IRIS^{2+}}$). This comparison refers to the inversion considering all the \irissqp\ lines with the same weights both in the STiC and the \irissqp\ inversions. 
The IRL$_{STiC}$ and IRL$_{IRIS^{2+}}$ maps (top row) are visually equivalent, and most of the discrepancies are due to outliers, i. e. pixels associated with a bad inversion or with a bad calculation of the radiative loss, which are usually related. There is a region where the results are not correct. These locations are around [X, Y]=[-305, 110] and [X, Y]=[-275, 75]. An episodic event (not necessarily a flare) occurs in these locations. Likely, the profiles in these regions need a more sophisticated inversion scheme than the one used by STiC and \irissqp\ to be properly fitted. In addition to this region, there are also some scattered pixels, especially in the IRL$_{STiC}$ map, that show larger values than their surroundings. This is usually caused by a problem during the inversion and, therefore, in the calculation of the radiative losses associated with the atmosphere resulting from that inversion. As already mentioned, for some of these pixels \irissqp\ finds a solution more in line with the surrounding pixels than STiC does, and therefore the IRL from \irissqp\ looks better in these pixels. 
The frequency distribution for both inversions (left panel, bottom row) looks very similar; both have a mode $\approx~-4.4~$ kW~m$^{-2}$. This value is mostly located in the quiet Sun areas in the map, and it is similar to the value found in the literature for the quiet Sun \citep{Withbroe77, Vernazza81, DiazBaso21}. The scatter plot (right panel, bottom row) shows a strong linear correlation ($r_p \approx 0.9$) after filtering outliers. 

We have evaluated the linear fit considering the original values (blue line), the values filtered with the Z-score method (orange dashed line), and with the inter-quartile range (IRQ) method (green dashed line). The $Z$-score for a value $x$ in a data distribution is defined as $(x-\mu)/\sigma$, being $\mu$ the mean and $\sigma$ the standard deviation. The Z-score method discards those x values above or below a given $Z$, Thus, in this study, we have chosen a $Z$-score = $\pm3$, that is, those $x$ larger (+) or smaller (-) than $\mu \pm\times3\sigma$ are discarded. The IRQ method discards the values above $Q3+1.5\times IQR$ or below $Q1-1.5\times IQR$, being Q1 and Q3 the first (25$^{th}$) and the third (75$^{th}$) percentile, respectively, and the IQR is $Q3-Q1$. These outlier detection methods have also been applied to other linear correlations computed in the rest of this paper\footnote{A linear regression should be formally written as $\widehat{y} = a+bx$, where $\widehat{y}$ denotes the predicted value for y. For the sake of clarity in the typography, we have denoted the prediction of the dependent variable in the linear regression fit without the wide hat symbol.}.

\begin{figure*}
    \centering
    \includegraphics[width=1\linewidth]
    {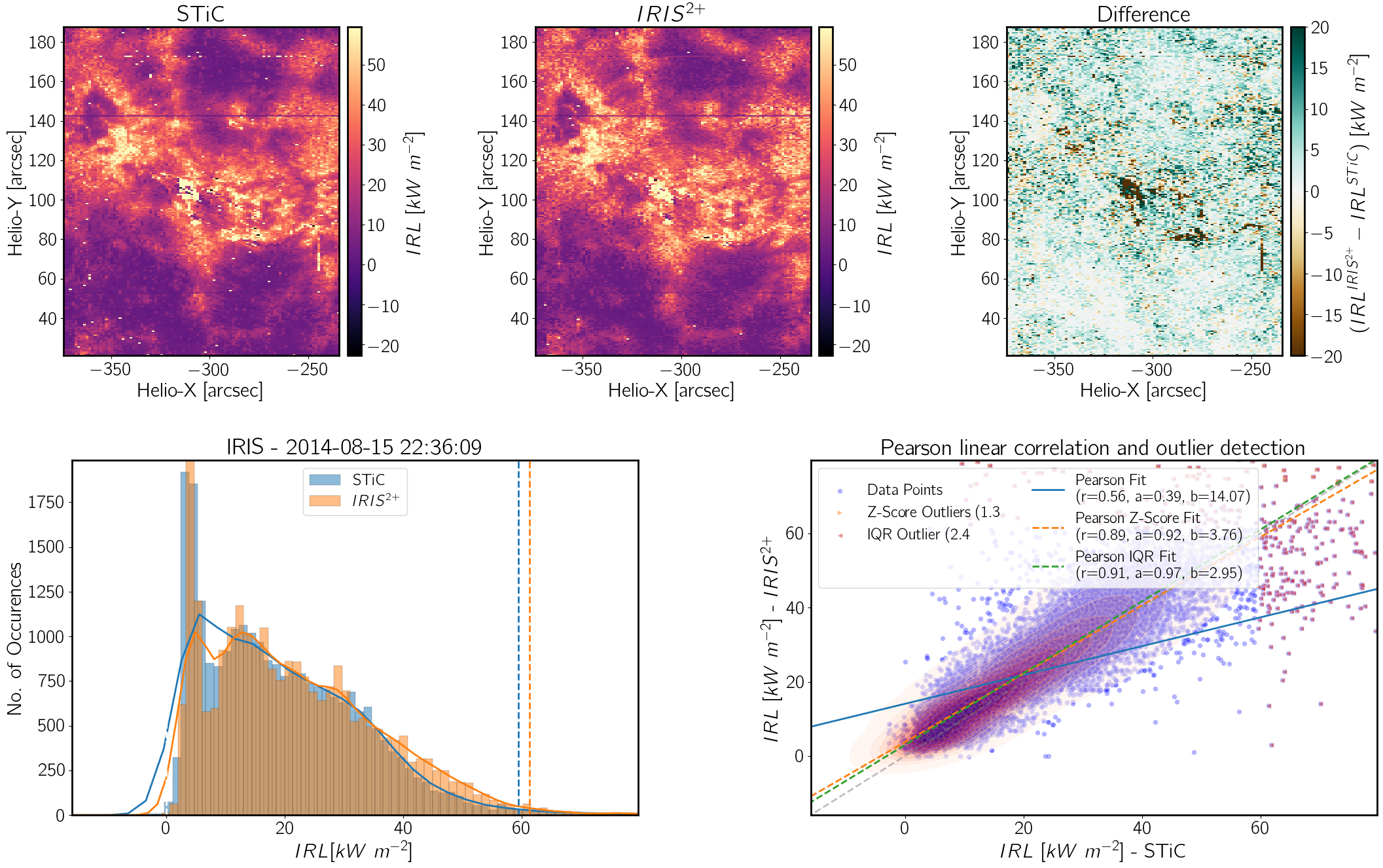}
    \caption{Comparison between the integrated radiative loss obtained by STiC (IRL$_{STiC}$) and by \irissqp\ (IRL$_{IRIS^{2+}}$).}
    \label{fig:lincorr_irl}
\end{figure*}

The results for other experiments that consider different selections of \irissqp\ lines (see Appendix  \ref{appendix:detailed_comparison_chi2} for details) are summarized in Figure \ref{fig:all_lincorr_irl}. After filtering the outliers, all the experiments except the ``{\it \cii}'' and the ``{\it Photospheric}'' show a strong linear correlation ($r_P\approx~0.75$) for all the data sets. The fact that the ``{\it Photospheric}'' experiments (magenta line) are providing the worst correlation is not surprising, since in this case the \irissqp\ is only fitting the photospheric lines, while STiC results always consider all the lines. Therefore, the information recovered in the chromosphere by \irissqp\ from this experiment has a large uncertainty and should not be considered as viable. For the ``{\it \cii}'' experiment (purple line), the linear correlation is moderate in some data sets, but weak in others. This tells us that recovering the information from the mid and high chromosphere using the \cii\ lines only is not enough to get a reasonable value of the radiative losses. In fact, while the correlation coefficient for the ``{\it \mgiik}'' experiment (brown line) is strong, the slope in some of the data is close to 0.5. In conclusion, we can deduce that we should always try to combine the \cii\ lines, the \mgii\ lines, and the \mguv\ lines to get a reasonable IRL. This is not difficult, since some combination of these lines is always present in any IRIS observation.\footnote{See Table 1 in \cite{SainzDalda24} for the IRIS linelists in the IRIS NUV channel.}

In summary, the IRL obtained by \irissqp\ and the one obtained by STiC overall agree quite well. 

\begin{figure*}
    \centering
    \includegraphics[width=1\linewidth]
    {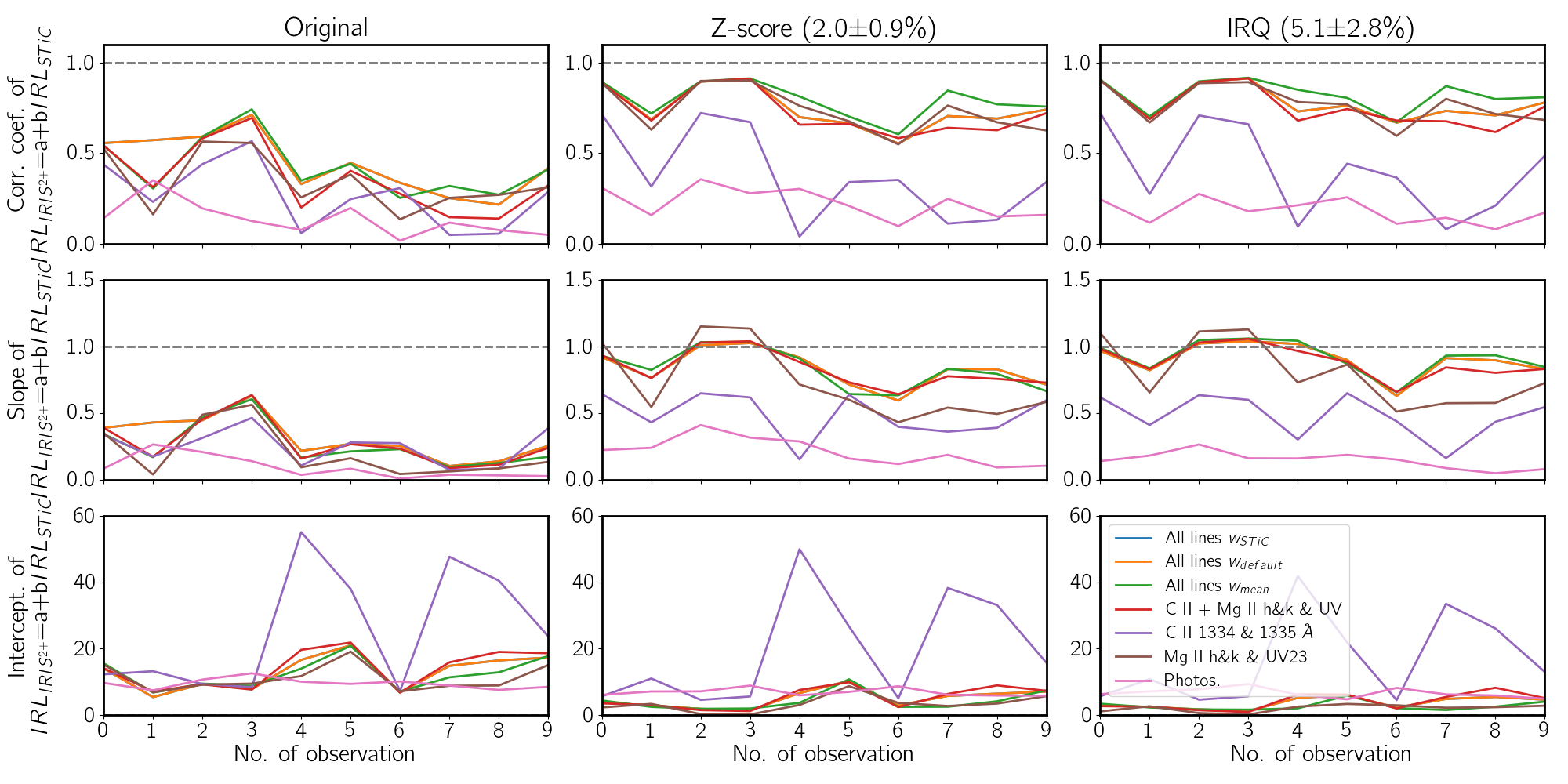}
    \caption{Linear correlation fit parameters for IRL$_{IRIS^{2+}} = a+bIRL_{STiC}$ for all the original data (first column), the Z-score filtered data (second column), and the IQR filtered data (third column).}
    \label{fig:all_lincorr_irl}
\end{figure*}

\subsection{Comparison of the thermodynamic results obtained by STiC and \irissqp}\label{sec:comparison_thermo}

We have followed a similar strategy in this section as in the previous ones: first, we visualize the frequency distribution of the thermodynamic values at each optical depth obtained from each inversion method, a linear correlation between them, and the map showing the values of that thermodynamic variable obtained by STiC and \irissqp\ at the given optical depth. We have done this for the temperature ($T$), \vlos, and \vturb\ obtained by STiC and \irissqp. As before, we have performed this linear fit between the unfiltered results obtained by STiC and by \irissqp, but now we have also performed the linear fit using the filter methods mentioned above to reduce the influence of outliers. We have not compared the \nne, and other thermodynamic variables derived from the temperature. 

Figures \ref{fig:lincorr_T_1} and \ref{fig:lincorr_T_2} shows the comparison between the temperature (in k$K$) at optical depths \ltau equal to -6, -4, -3, and -1, as obtained by STiC and by \irissqp. The left panel shows the frequency distribution of the temperature for each inversion method. The middle panel shows the scatter plot of these variables and the fits for the linear correlation of the original data (blue line), the filtered data with  Z-score method (orange dashed line), and the IQR method (green dashed line). The panels on the right side show the temperature maps at the optical depth, as obtained by STiC and \irissqp. These figures show some similarity with what we found in the comparison of different weighting methods and line selections discussed in Appendix \ref{appendix:detailed_comparison_chi2}. 
The slope and the intercept value suggest a discrepancy between T$_{STiC}$ and T$_{IRIS^{2+}}$. At \ltau=-6 (first row), the frequency distribution T$_{STiC}$ and T$_{IRIS^{2+}}$ is similar, but the peak (the mode) of the distribution of T$_{IRIS^{2+}}$ is slightly shifted to the left, i.e., it is cooler. These two behaviors are actually also reflected in the map (despite the visual appearance): in the T$_{STiC}$ map some areas are hotter and some that are cooler than in the T$_{IRIS^{2+}}$ map, but in general, the T$_{IRIS^{2+}}$ is slightly cooler than the T$_{STiC}$ map. Therefore, in detail, the maps, the scatter plot, the fits, and the frequency distribution tell the same story.

\begin{figure*}
    \centering
    \includegraphics[width=.9\linewidth]
    {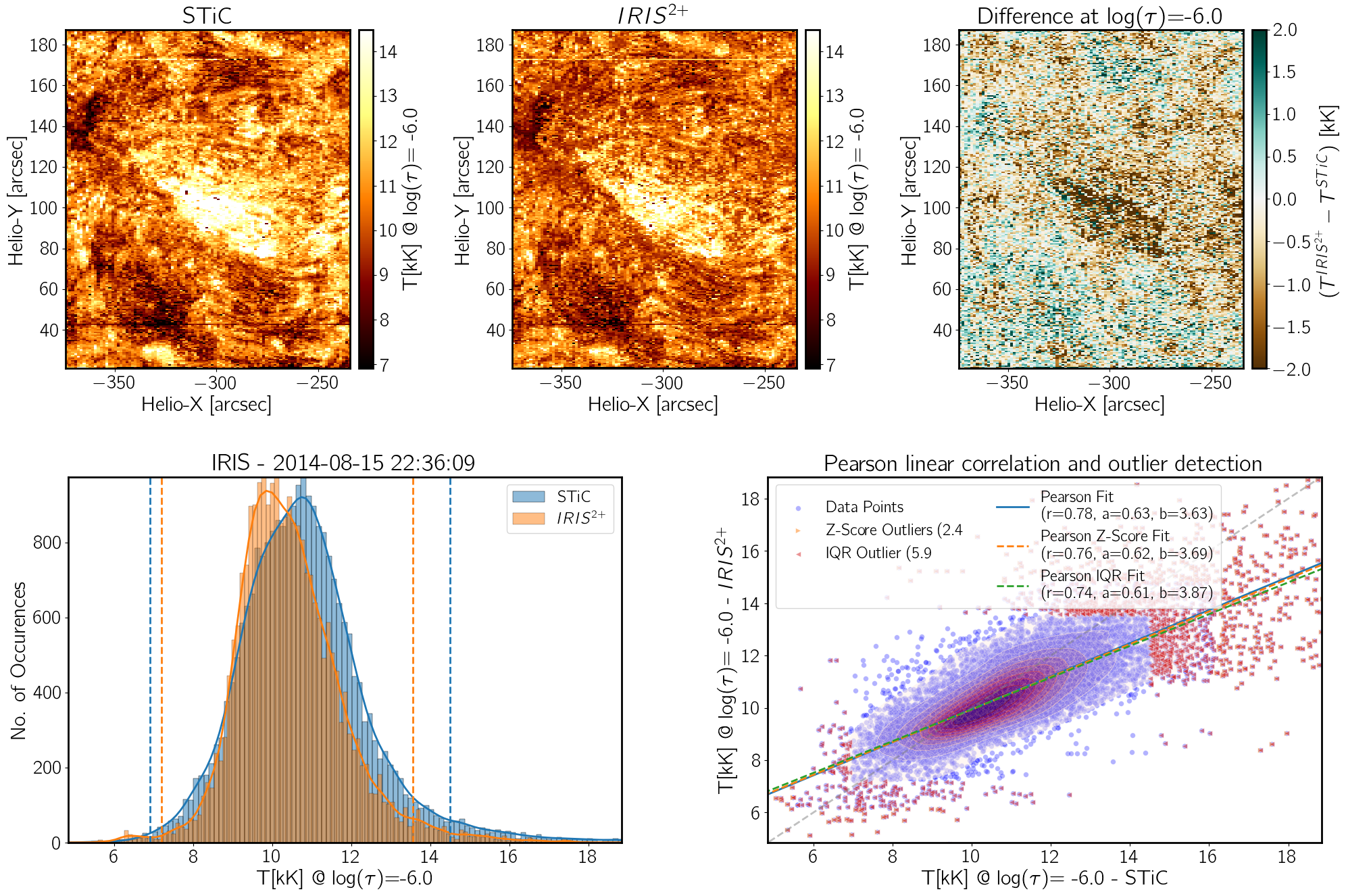}
    \includegraphics[width=.9\linewidth]
    {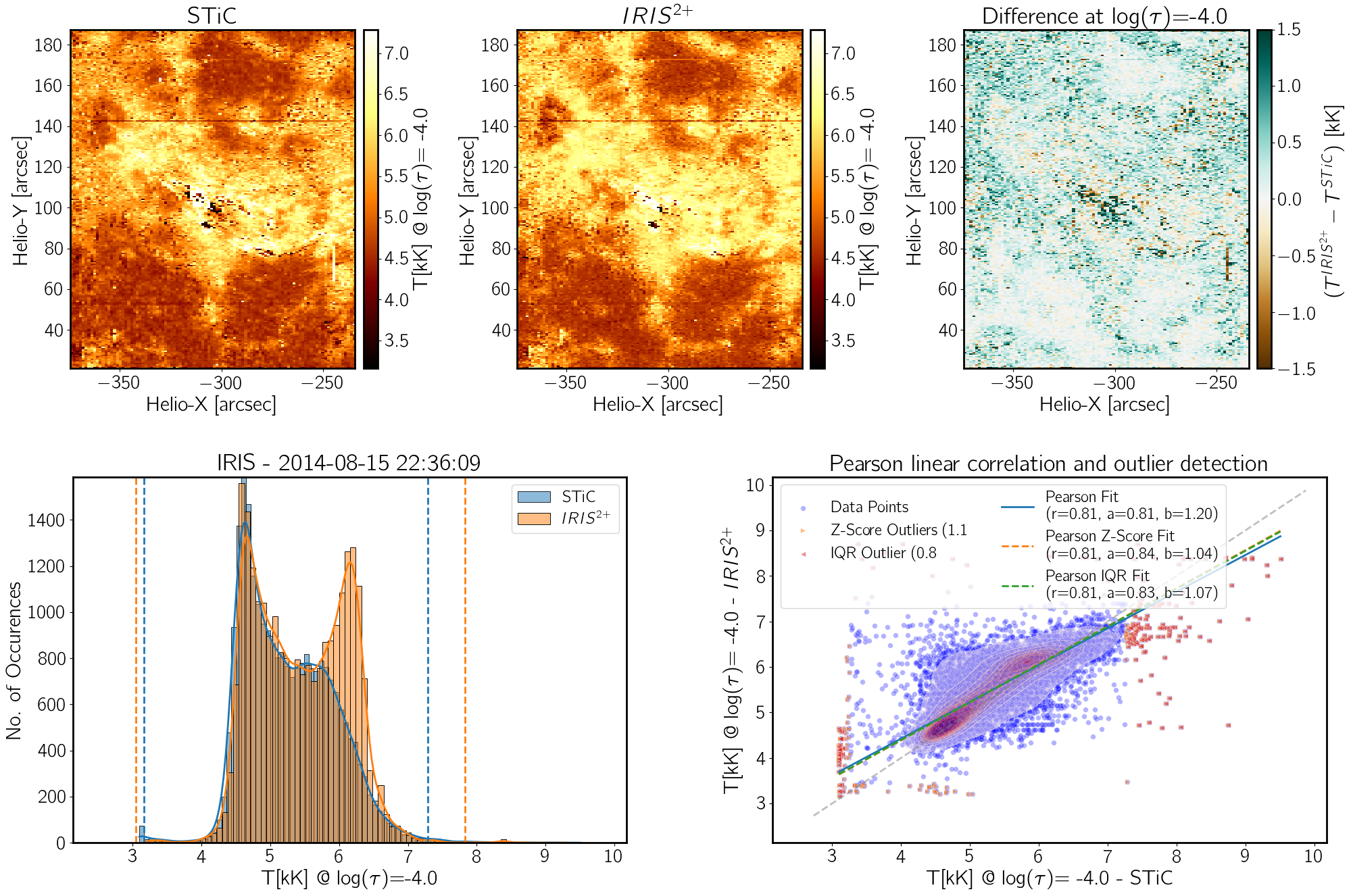}
   \caption{Comparison between the T (in k$K$) obtained by STiC and by \irissqp\ at the high chromosphere (top panels, \ltau=-6) and the mid chromosphere (bottom panels, \ltau=-4). See the section \ref{sec:comparison_thermo} for more details.}\label{fig:lincorr_T_1}
\end{figure*}  
\begin{figure*}
    \centering
    \includegraphics[width=.9\linewidth]
    {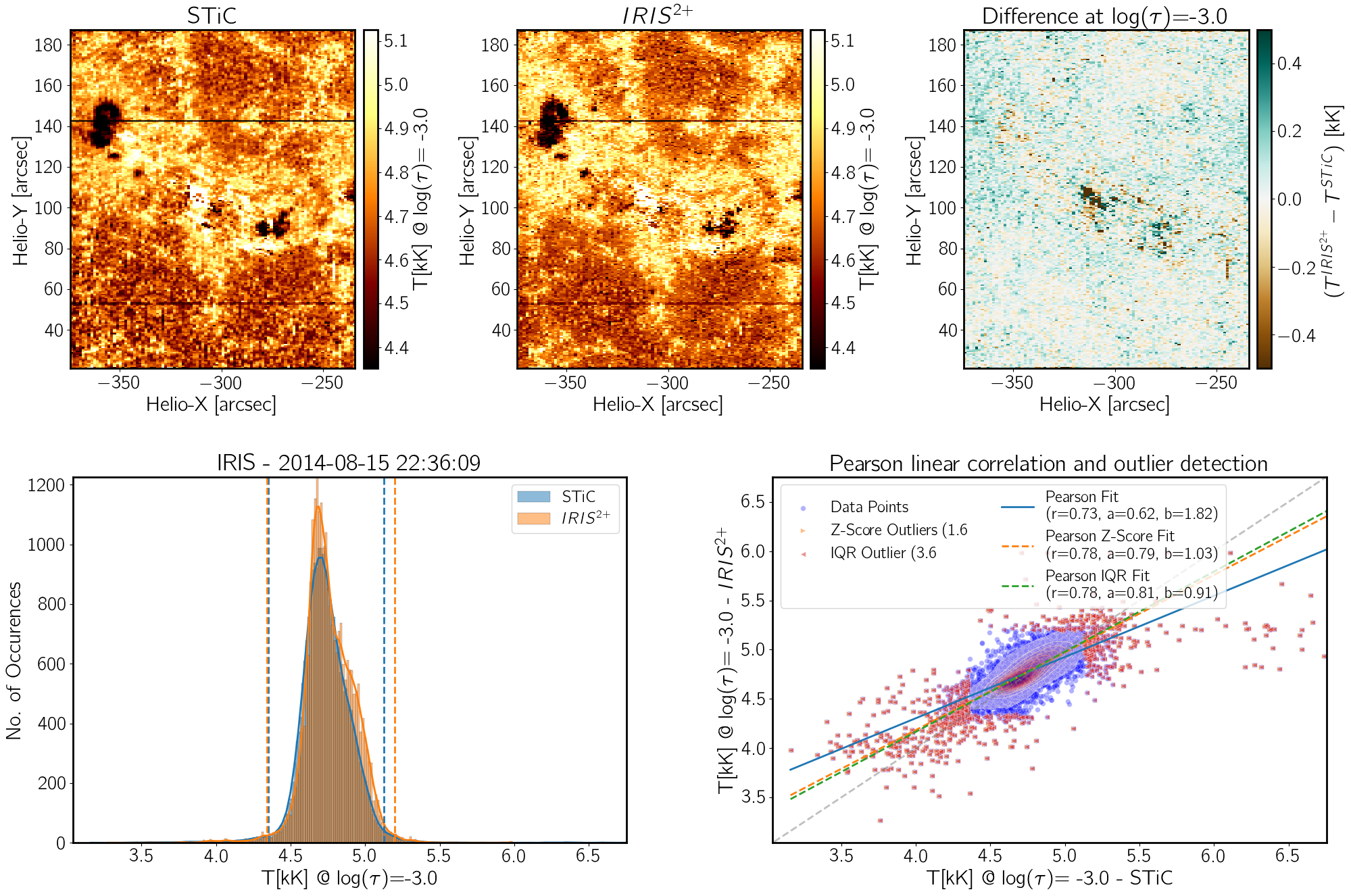}
    \includegraphics[width=.9\linewidth]
    {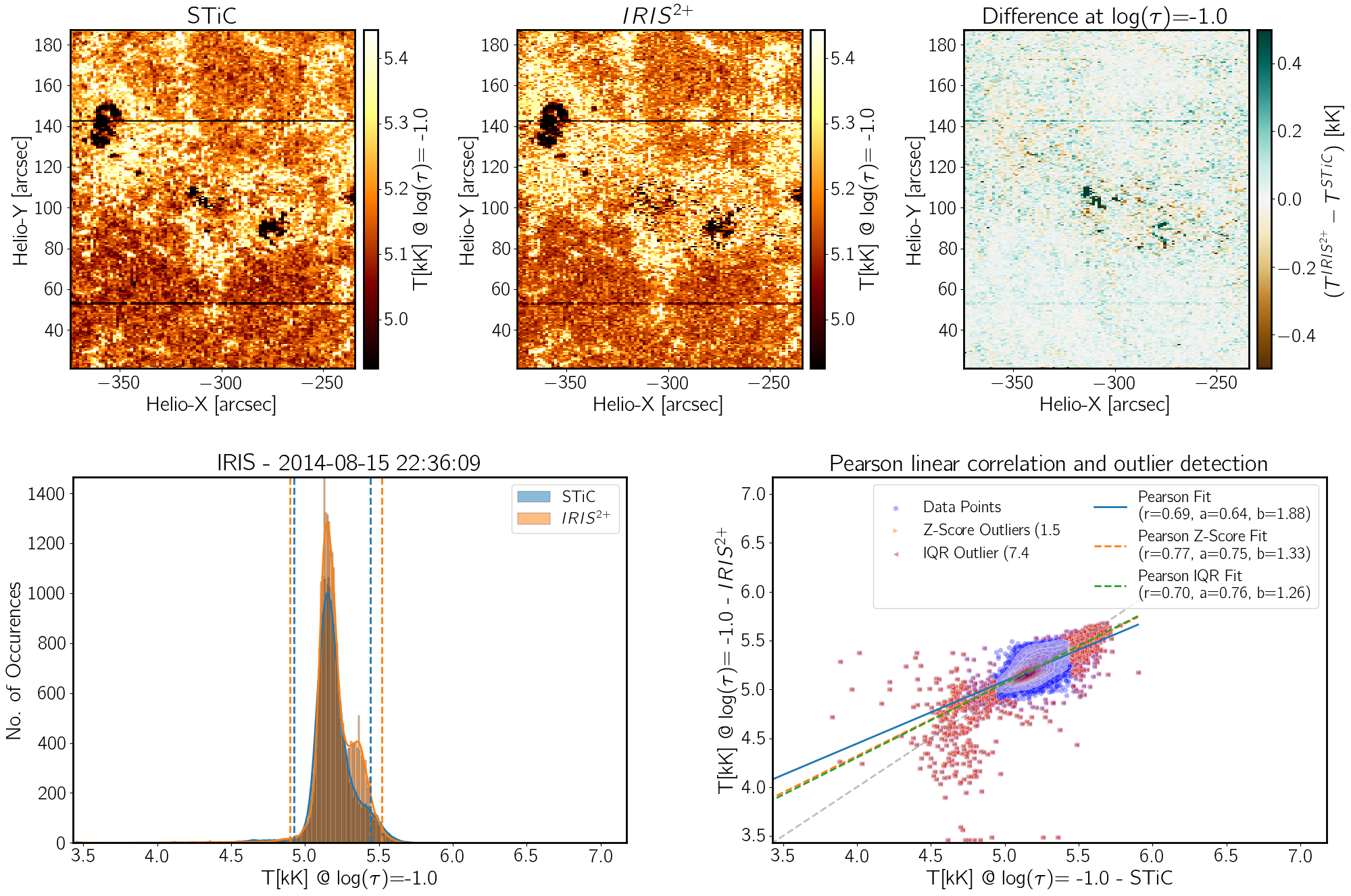}
   \caption{Comparison between the T (in k$K$) obtained by STiC and by \irissqp\ at the low chromosphere (top panels, \ltau=-3) and mid photosphere (bottom panels, \ltau=-1). See the section \ref{sec:comparison_thermo} for more details.}\label{fig:lincorr_T_2}
\end{figure*}

%
The frequency distribution plots for the temperature in Figure \ref{fig:lincorr_T_2} show
a behavior common for many of the data sets analyzed in this study. At some optical depths, the distribution corresponding to the $T_{IRIS^{2+}}$ (in orange) shows a double peak, usually one is mostly matching the more dominant distribution of $T_{STiC}$, while the other peak is lower in the number of occurrences and located at higher temperatures. This can be seen in the frequency distribution for 
T$_{IRIS^{2+}}$ at \ltau = -4 (first panel in the second row of Figure \ref{fig:lincorr_T_1}) and at
\ltau = -1 (first panel in the bottom row of Figure \ref{fig:lincorr_T_2}). As a consequence, \irissqp\ tends to provide a hotter solution than STiC does. This discrepancy will be discussed in Section \ref{sec:discussion}.

The frequency distribution for the values obtained by \irissqp\ is often narrower and slightly more peaked than the frequency distribution for the STiC inversion. This is especially noticeable in the case of \vlos. Figures \ref{fig:lincorr_vlos_1} and \ref{fig:lincorr_vlos_2} show the comparison between \vlos$_{STIC}$ and \vlos$_{IRIS^{2+}}$. For \ltau\ = -6 (first row in Figure \ref{fig:lincorr_vlos_1}) the linear correlation between these values is strong (the Pearson correlation coefficient, $r_p$, is $\approx 0.7$). We also appreciate this correlation in the frequency distribution and the maps. As we will see later, this strong correlation happens for $-7\lessapprox$\ltau$\lessapprox~-5$, i.e., in the high chromosphere. Despite the apparent visual similarity in some of the \vlos\ maps shown in this figure, the linear correlation for optical depths $-4\lessapprox$\ltau is low. The most striking case is for the \vlos\ maps at \ltau=-3 (top panels in Figure \ref{fig:lincorr_vlos_2}, which look very similar to each other, while the $r_p\approx0.01$, which means there is no linear correlation between \vlos$_{STIC}$ and \vlos$_{IRIS^{2+}}$. 

\begin{figure*}
    \centering    
    \includegraphics[width=.9\linewidth]
    {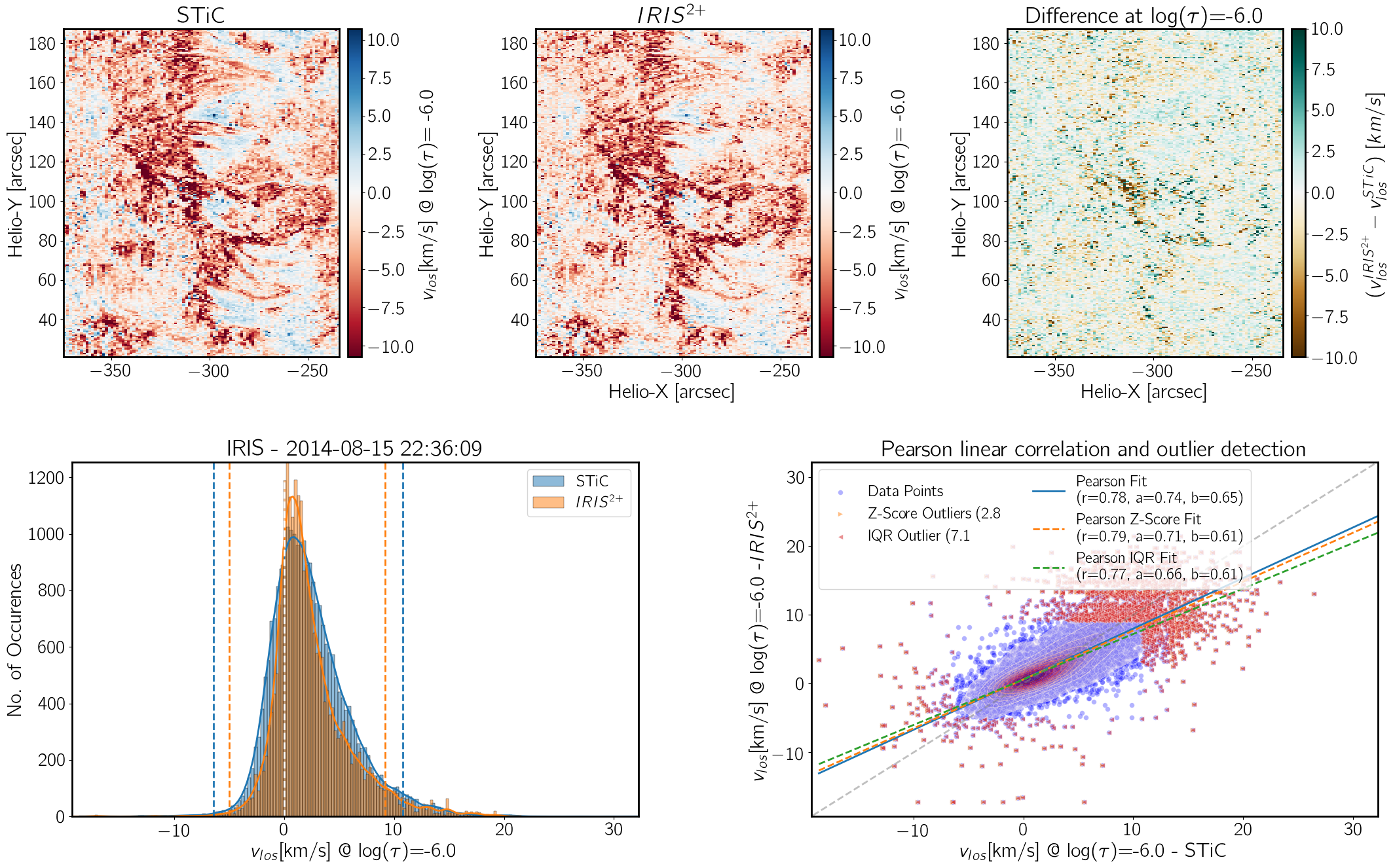}
    \includegraphics[width=.9\linewidth]
    {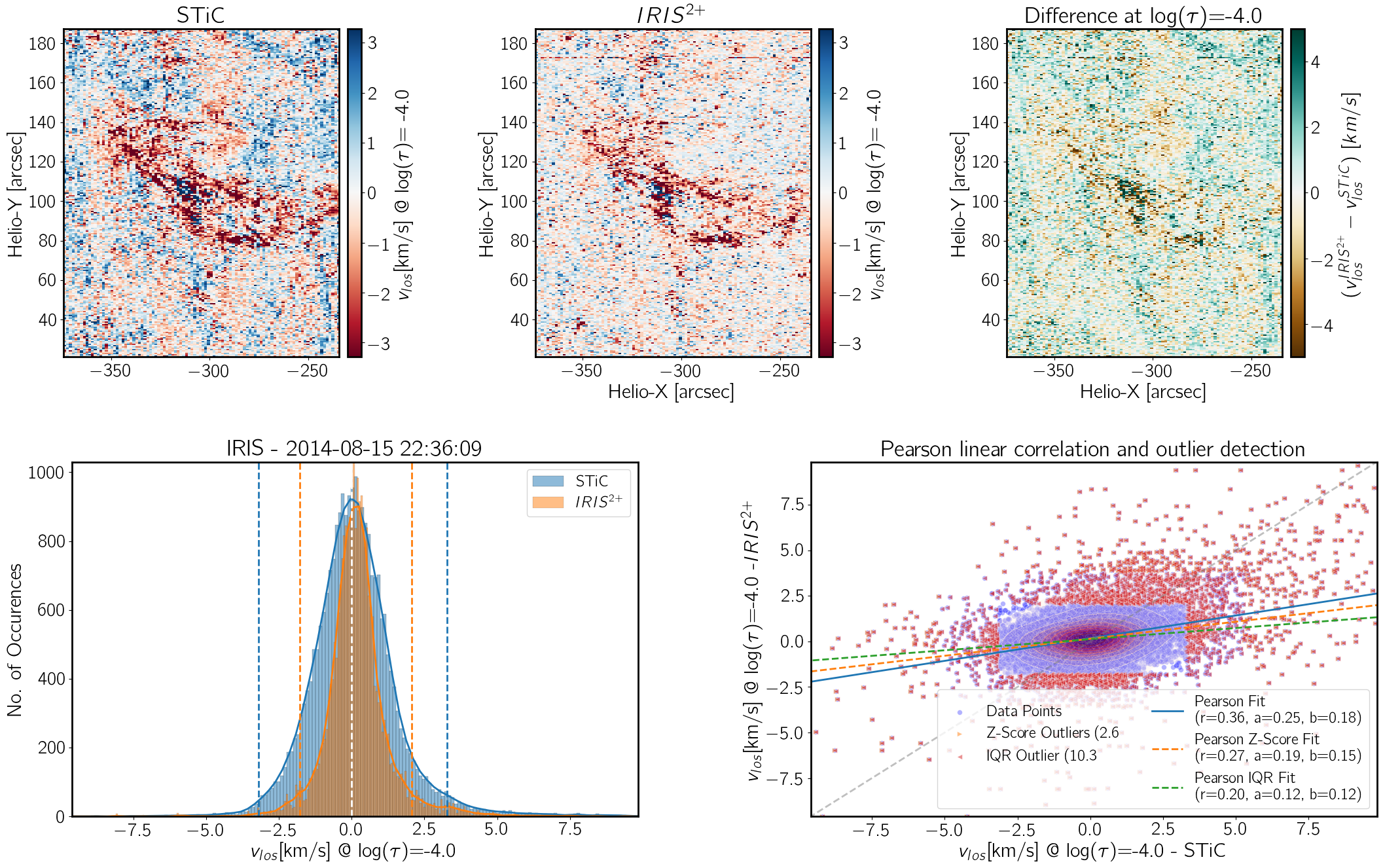}
    \caption{Comparison between the \vlos\ (in \kms) obtained by STiC and by \irissqp\ at the high chromosphere (top panels, \ltau=-6) and the mid chromosphere (bottom panels, \ltau=-4). See the section \ref{sec:comparison_thermo} for more details.}\label{fig:lincorr_vlos_1}
\end{figure*}

\begin{figure*}
    \centering    
    \includegraphics[width=.9\linewidth]
    {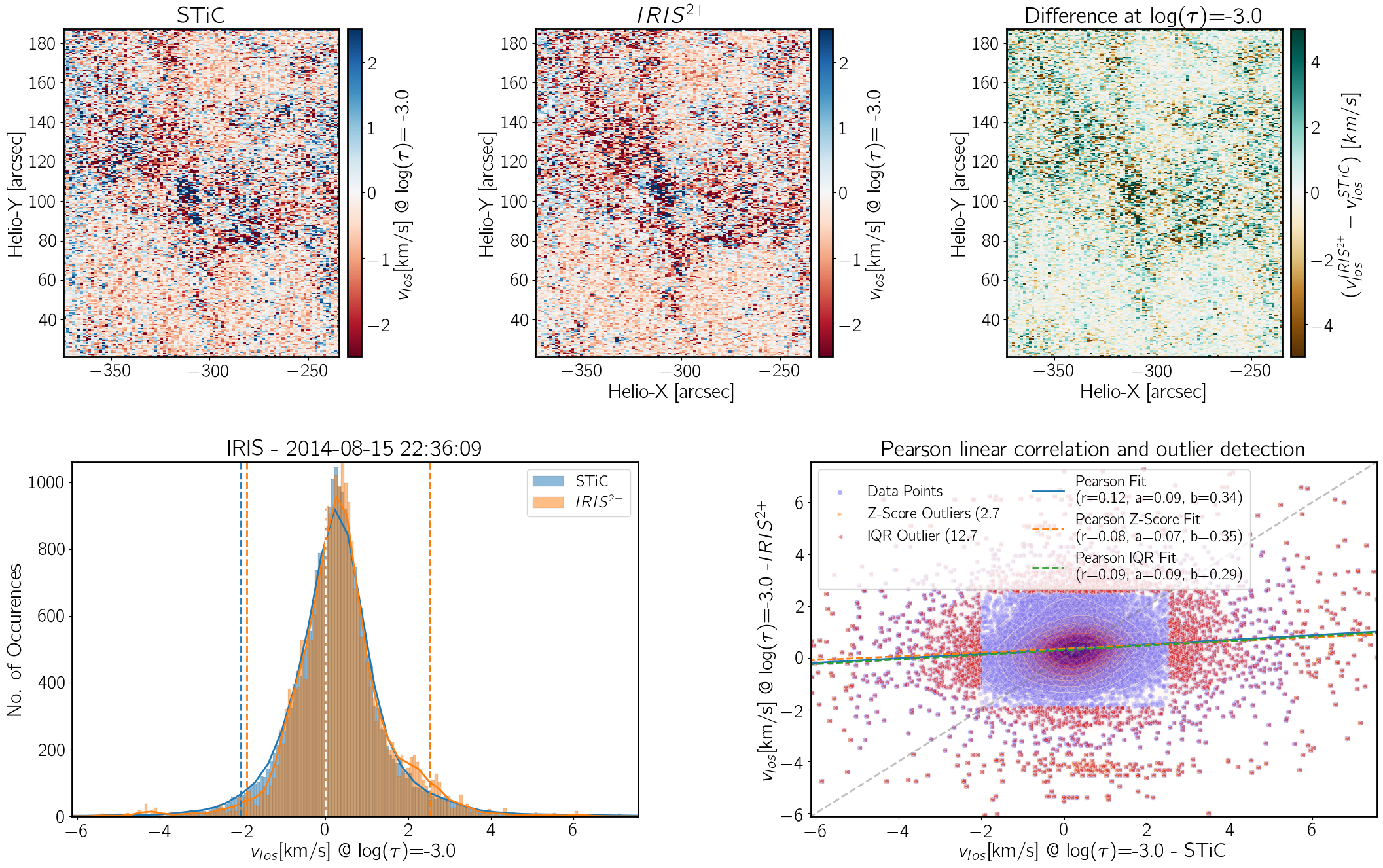}
    \includegraphics[width=.9\linewidth]
    {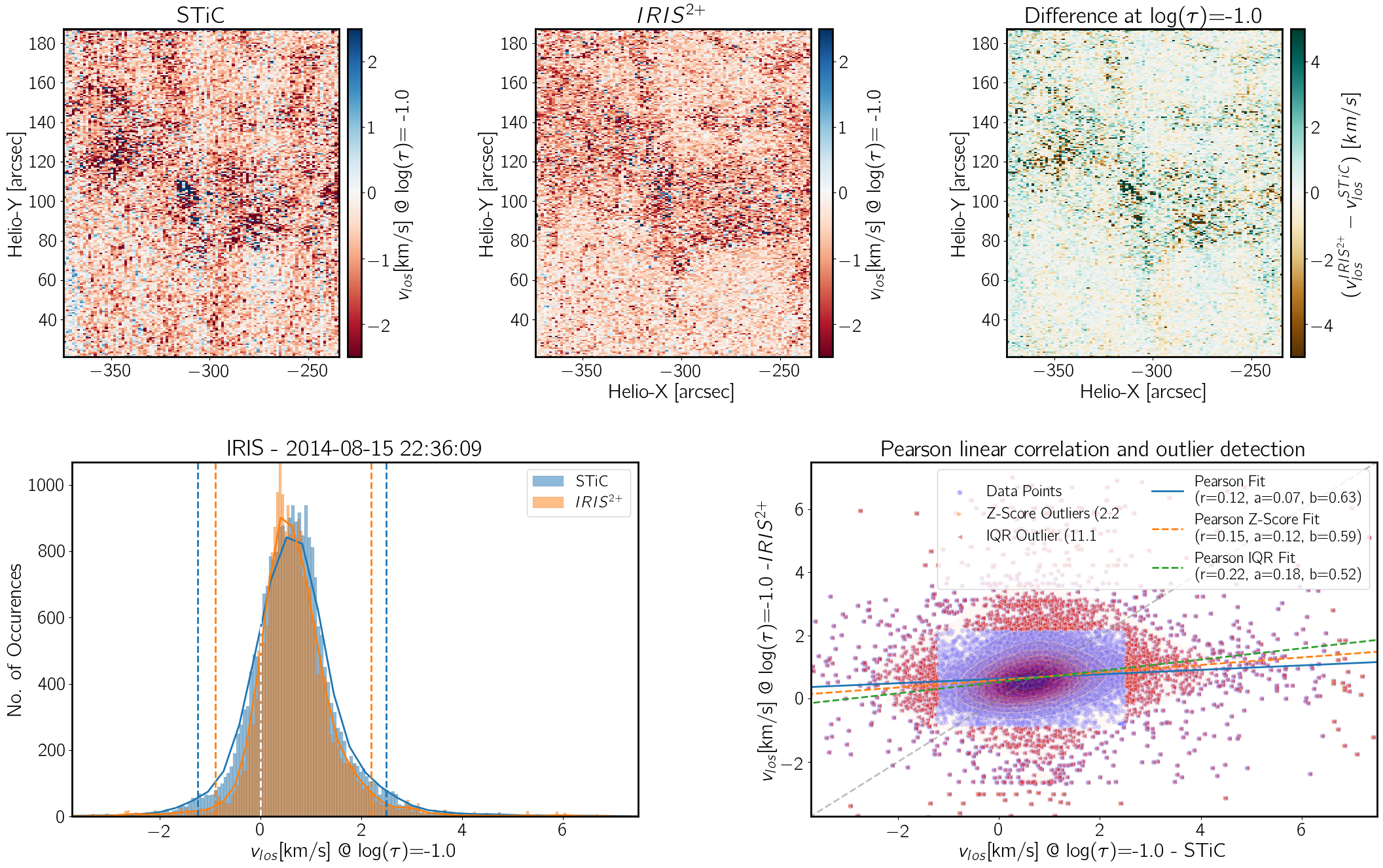}
    \caption{Comparison between the \vlos\ (in \kms) obtained by STiC and by \irissqp\ at the low chromosphere (top panels, \ltau=-3) and mid photosphere (bottom panels, \ltau=-1). See the section \ref{sec:comparison_thermo} for more details.}\label{fig:lincorr_vlos_2}
\end{figure*}

Figures \ref{fig:lincorr_vturb_1} and \ref{fig:lincorr_vturb_2} shows the comparison for the \vturb\ obtained by STiC and \irissqp. The linear correlation between them is strong both at \ltau=-6 and \ltau=-4, with $r_p \gtrapprox0.7$, weak at \ltau=-3, and negligible at \ltau=-1.
Again, the frequency distribution of \vturb$_{IRIS^{2+}}$ is narrower than the one corresponding to \vturb$_{STiC}$.

All the previous figures were for the first observation (\#0) of Figure \ref{fig:intmap_1}. To summarize the behavior of all the data sets that we have considered in this investigation, we show in Figure \ref{fig:lincorr_thermo_3x3} the correlation coefficient, slope, and intercept for a linear fit of T, \vlos, and the \vturb\ for all the datasets shown in Figures \ref{fig:intmap_1} and \ref{fig:intmap_2}. The colors of the lines in the plots of Figure \ref{fig:lincorr_thermo_3x3} correspond to the color of the labels of each dataset in Figures \ref{fig:intmap_1} and \ref{fig:intmap_2}.

For the temperature, the correlation is between moderate and strong ($r_P \gtrapprox~0.5$) for most of the data sets between $-7\lessapprox$\ltau$\lessapprox~-1$, with a moderate correlation at \ltau=-5 and \ltau=-2.8. The latter is because the correlation is weak at these optical depths for data sets \#7 and \#8. The slope gets close to 1 between $-4\lessapprox$\ltau$\lessapprox~-3$ for some data sets. For the rest of the optical depths, the slope is around 0.6 on average. The intercept is not meaningful in this case, since the value T$_{STiC}~=~0~K$ is not observed in the solar atmosphere. It is shown only for consistency with the other panels.

For \vlos\, the correlation coefficient and slope show a moderate correlation between $-7\lessapprox$\ltau$\lessapprox~-5$, in the high chromosphere. After the data are filtered, there is also a weak correlation ($r_P \approx 0.15$) between $-1.4\lessapprox$\ltau$\lessapprox~-0.4$, that is, in the photosphere. 
The intercept shows values \vlos$<~2~$\kms on average for all data sets at all the optical depths except for the data set \#7 (gray line).

The correlation coefficient for \vturb\ is moderate to strong for most of the datasets for $-6\lessapprox$\ltau$\lessapprox~-3$. Again, the data set \#7 is the only one showing a moderate to weak correlation in these optical depths. Half of the datasets also show a moderate correlation at \ltau = -7.4, 
which is an optical depth where the \irissqp\ lines, especially the \cii\, may be sensitive to changes in the thermodynamics during some events, which have not been considered in this study or the \irissqp\ database. Nevertheless, we have to interpret this result carefully. The \vturb\ both in \irissq\ and \irissqp\ databases are artificially limited to 15 \kms, which is a reasonable limit for the \vturb\ in the chromosphere. This limit was imposed when these databases were built with STiC, and it is also imposed in the individual inversions with STiC in the data sets studied in this investigation. The correlation is, therefore, mostly between these high values, which dominate the frequency distribution at that optical depth. For the unfiltered values \vturb, the correlation is moderate in the photosphere, but the filtered data decreases this correlation to being weak to negligible in $-1\lessapprox$\ltau$\lessapprox~-0$. For the slope, the average value is $\approx 0.6$
at $-6\lessapprox$\ltau$\lessapprox~-4$, dropping to values below 0.2 at -3$<$\ltau. The intercept shows a high value of $\approx~-8~$\kms at \ltau=-7. At this optical depth, the correlation is moderate to weak and the slope is $\approx~0.2$. 

\begin{figure*}
    \centering    
    \includegraphics[width=.9\linewidth]
    {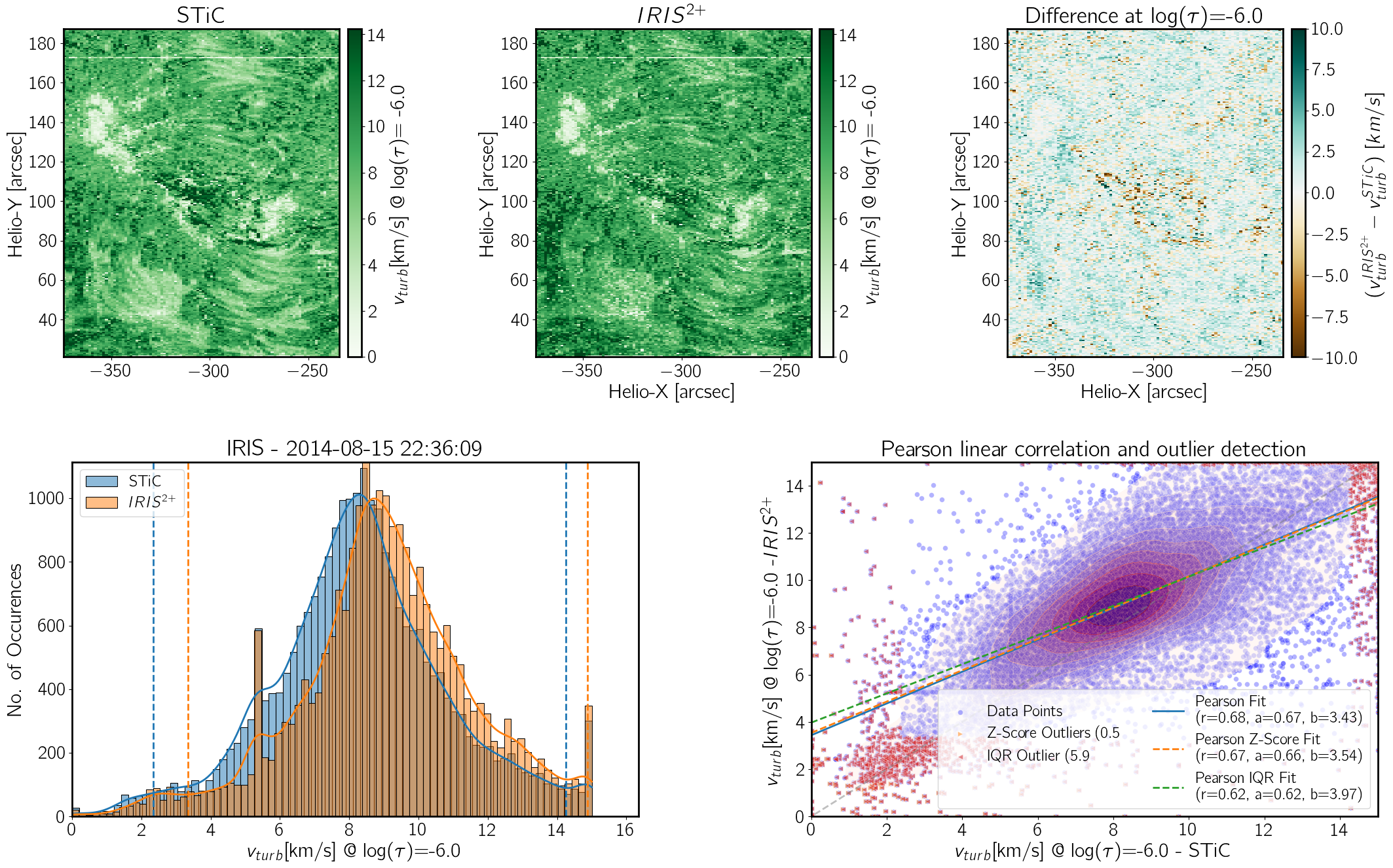}
    \includegraphics[width=.9\linewidth]
    {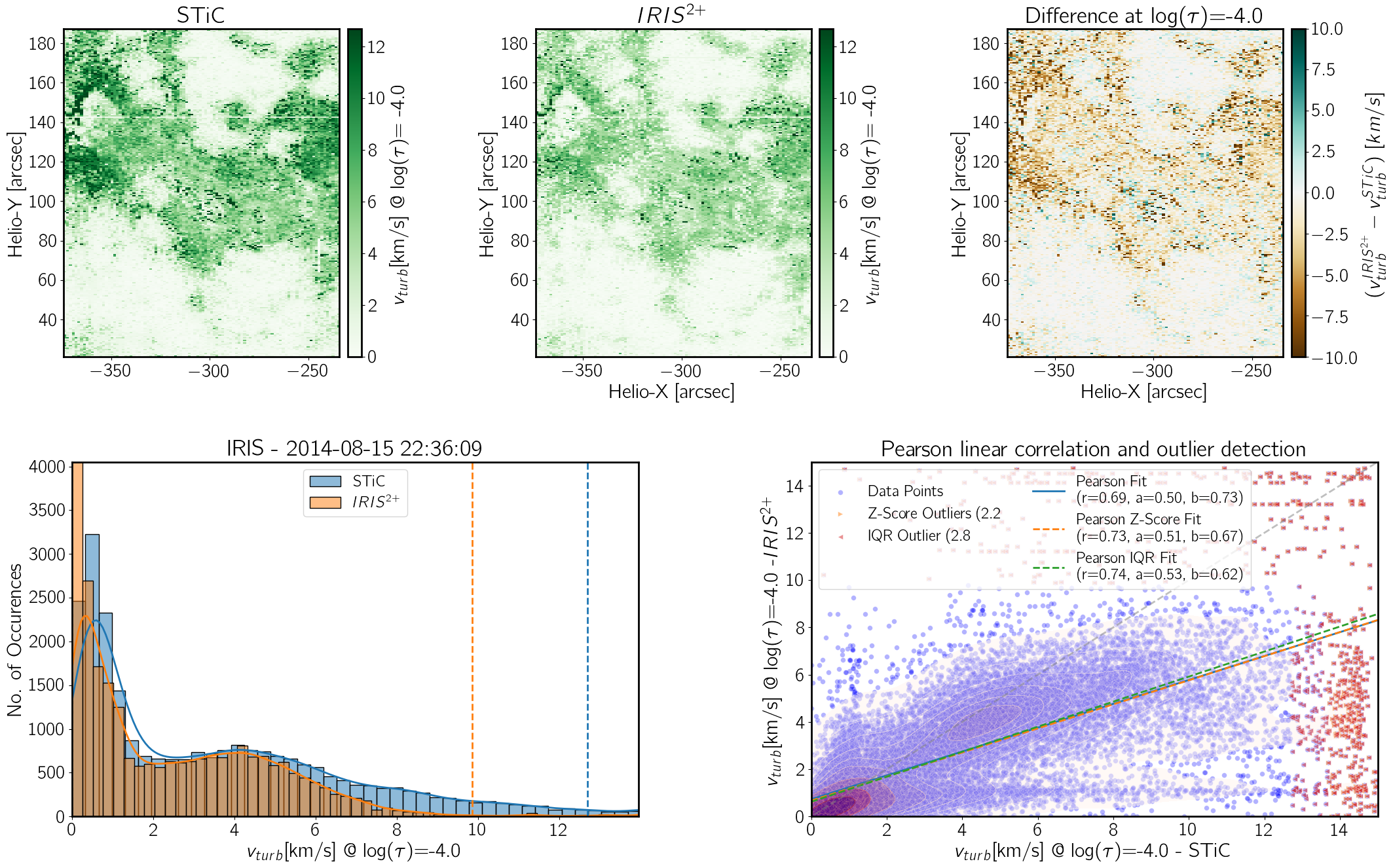}
    \caption{Comparison between the \vturb\ (in \kms) obtained by STiC and by \irissqp\ from the top of the chromosphere to the mid photosphere. See the section \ref{sec:comparison_thermo} for more details.}
    \label{fig:lincorr_vturb_1}
\end{figure*}

\begin{figure*}
    \centering    
    \includegraphics[width=.9\linewidth]
    {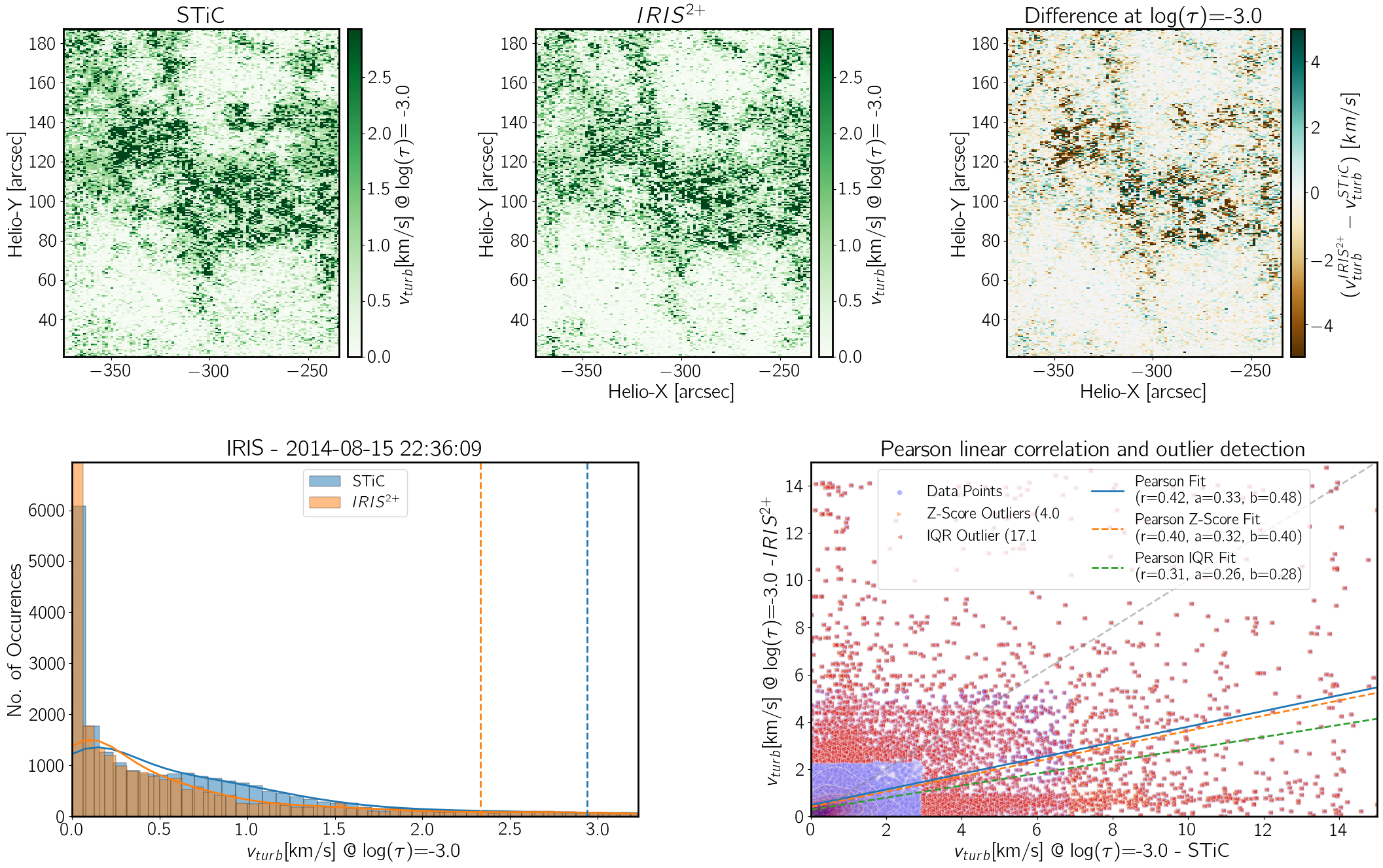}
    \includegraphics[width=.9\linewidth]
    {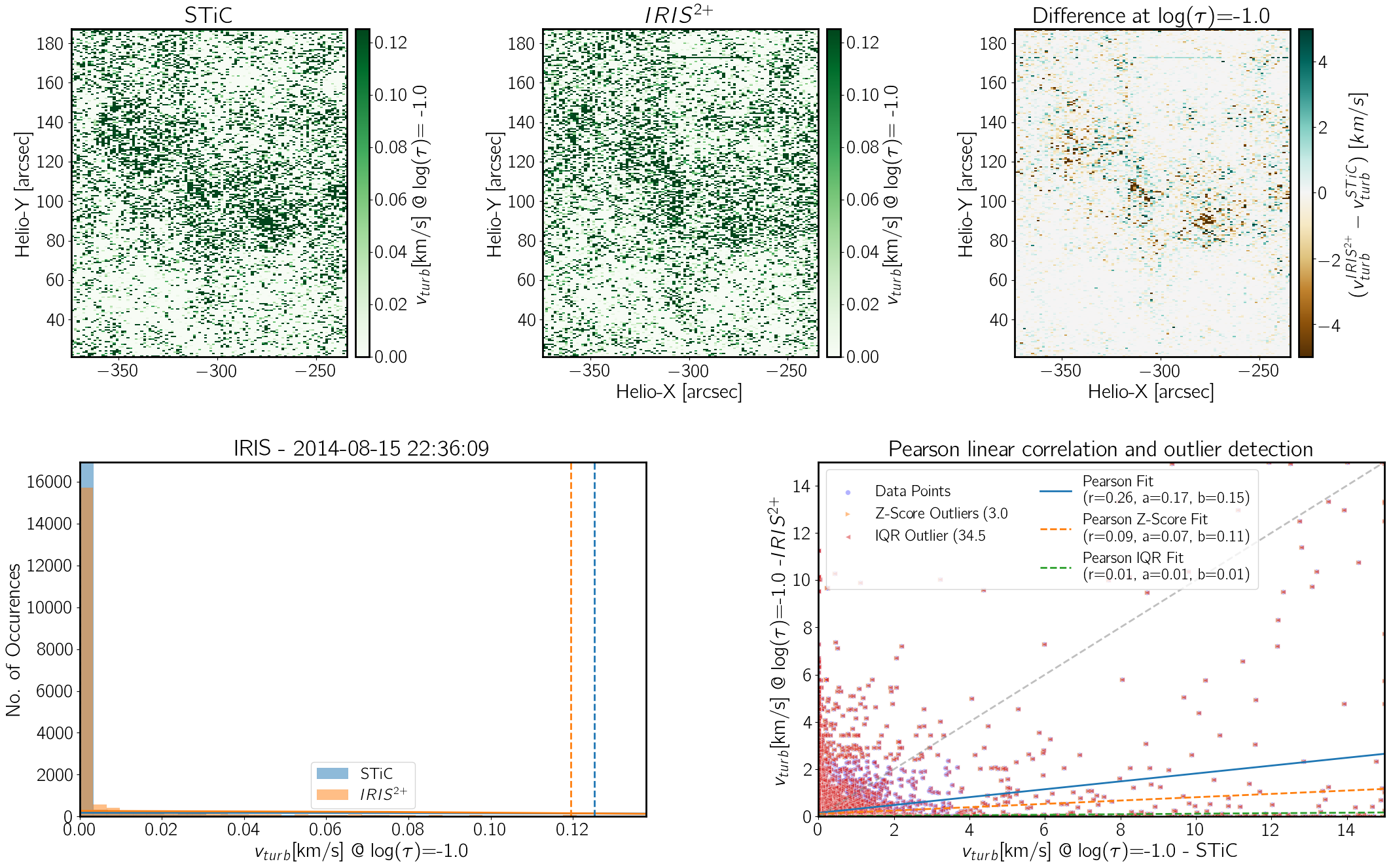}
    \caption{Comparison between the \vturb\ (in \kms) obtained by STiC and by \irissqp\ from the top of the chromosphere to the mid photosphere. See the section \ref{sec:comparison_thermo} for more details.}
    \label{fig:lincorr_vturb_2}
\end{figure*}

\begin{figure*}
    \centering
    \includegraphics[width=.80\linewidth]
    {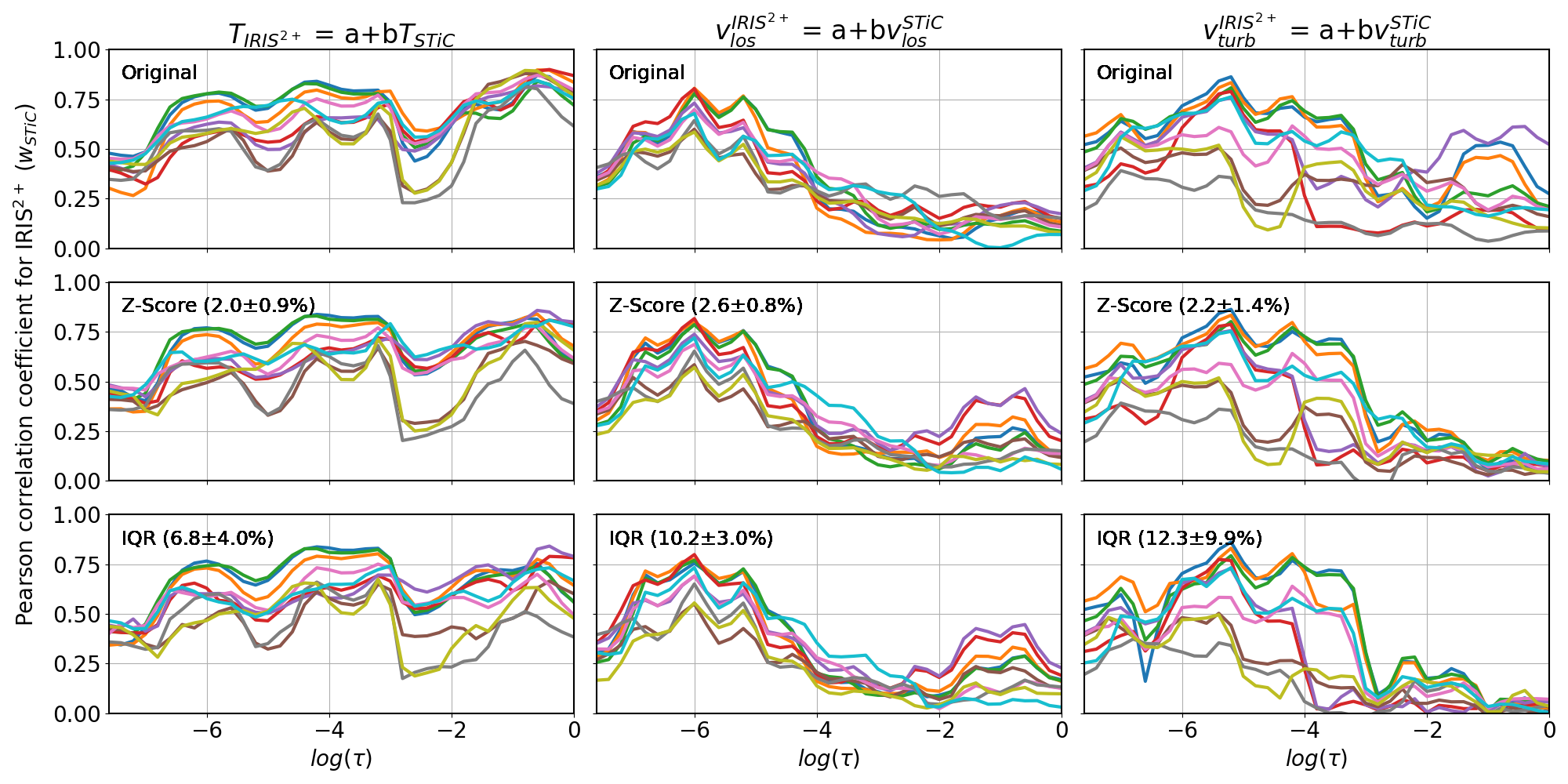}
    \includegraphics[width=.80\linewidth]
    {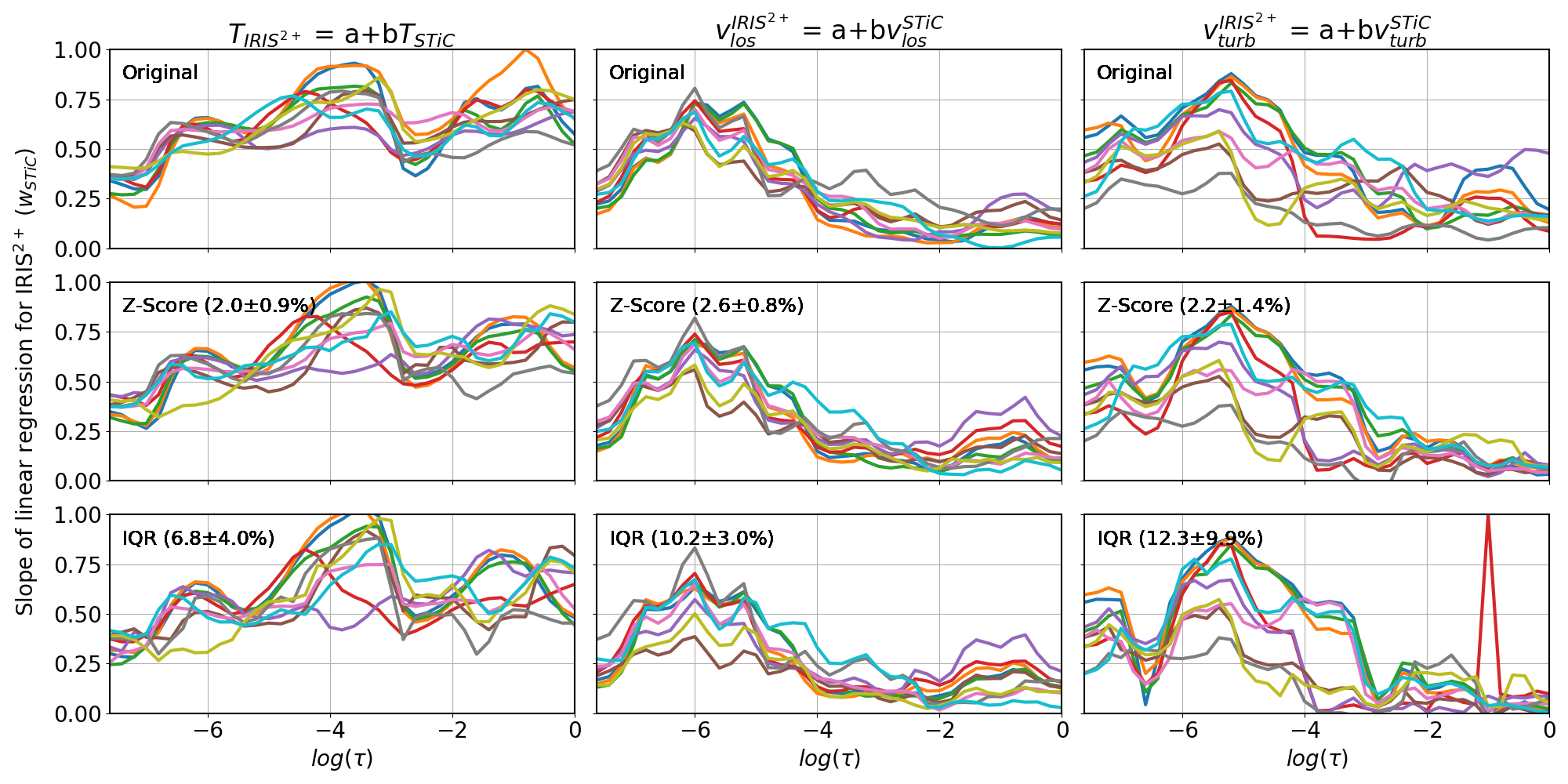}
    \includegraphics[width=.80\linewidth]
    {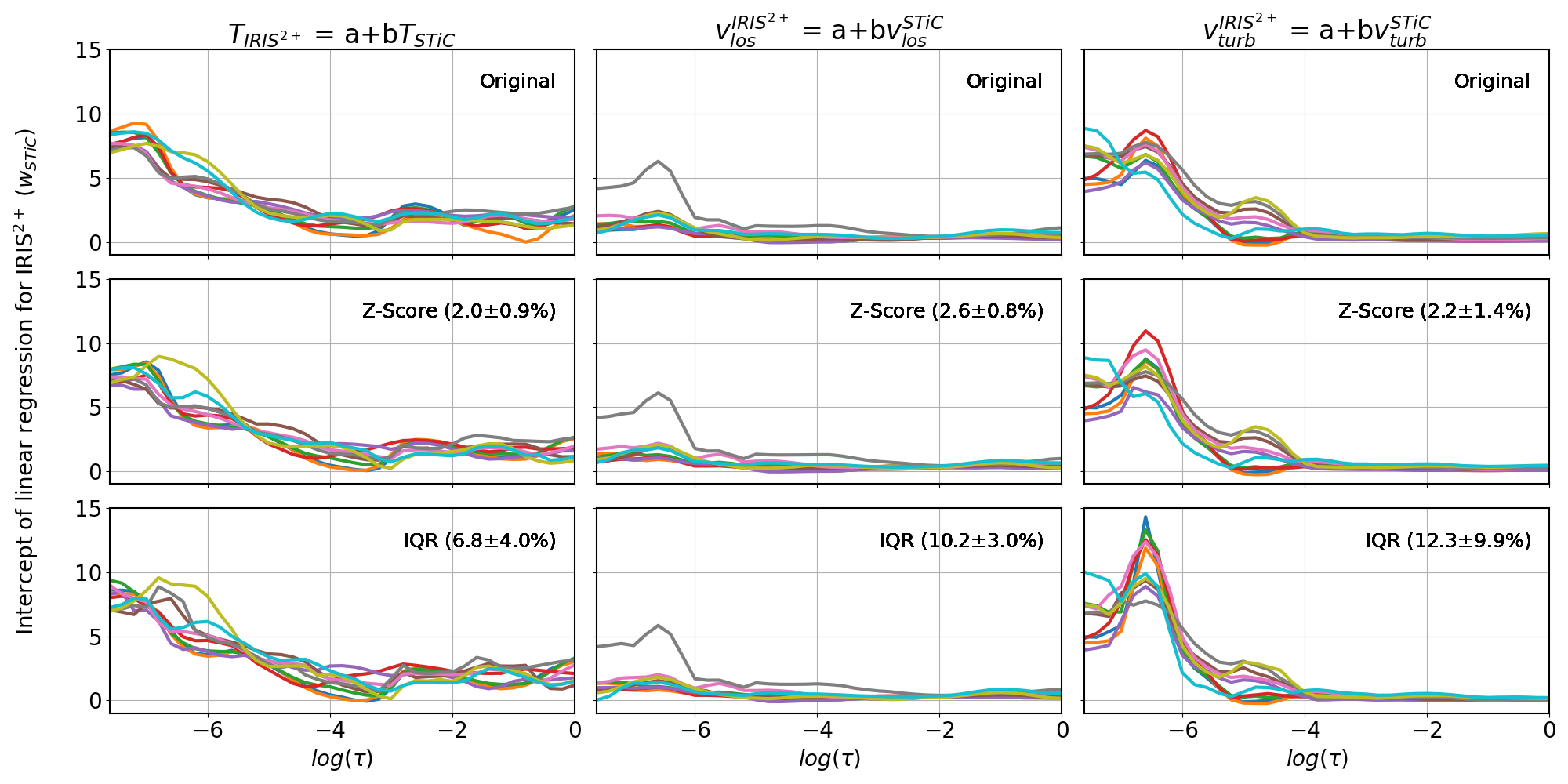}
    \caption{Pearson coefficient (top), slope (middle), and intercept (bottom) of the linear correlation between the thermodynamics values along the optical depth obtained by STiC and \irissqp. The color of the lines corresponds with the color of the label of each data set in Figures \ref{fig:intmap_1} and \ref{fig:intmap_2}.}
    \label{fig:lincorr_thermo_3x3}
\end{figure*}

\asd{Figure \ref{fig:plot_absdiff} shows, for two diffferent inversion tests, the spatially averaged differences between the IRL (and the thermodynamic parameters) obtained by \irissqp\ and STiC. To better interpret these differences, we have computed spatial averages that account for the sign of the difference, showing the positive difference in a solid line and the negative difference in a dashed line. The top panels show the results for the inversion test named {\it $w_{STiC}$} and the bottom panels for the case {\it $w_{mean}$}. The difference between these tests is the weights given to the lines (see Appendix \ref{appendix:detailed_comparison_chi2} for details). 
For the IRL, we see that, for both tests, for all the datasets used in this study, the spatially-averaged positive difference, i.e., IRL$^{IRIS^{2+}} >$ IRL$^{STiC}$, is $\approx$~6~kW~m$^2$, while the negative difference, i.e., IRL$^{IRIS^{2+}} <$ IRL$^{STiC}$, is $\approx$-10~kW~m$^2$. The absolute values of these differences are very close to each other, which indicates that, from a spatial average point of view, there is a similar overestimation and underestimation of IRL$^{IRIS^{2+}}$ with respect to IRL$^{STiC}$, and vice versa. The thermodynamic variables along the optical depth also show this behavior. Some datasets show a larger positive difference than a negative difference at -5 $<$ \ltau $<$ -3 in the top panels ($w_{STiC}$); however, this imbalance is lower for the bottom panels ($w_{mean}$), which means that this difference may be different depending on the weights given to the spectral lines during the inversions. 
We conclude that, for the datasets analyzed in this study, \irissqp\ may overestimate the thermodynamic parameters relative to STiC in some cases, while in others it may underestimate them. Statistically, however, the thermodynamic estimates obtained by both methods over an active region field of view are roughly consistent with each other.
}

\begin{figure*}
    \centering
        \includegraphics[width=0.8\linewidth,trim={10pt 150pt 10pt 150pt},clip]%
        {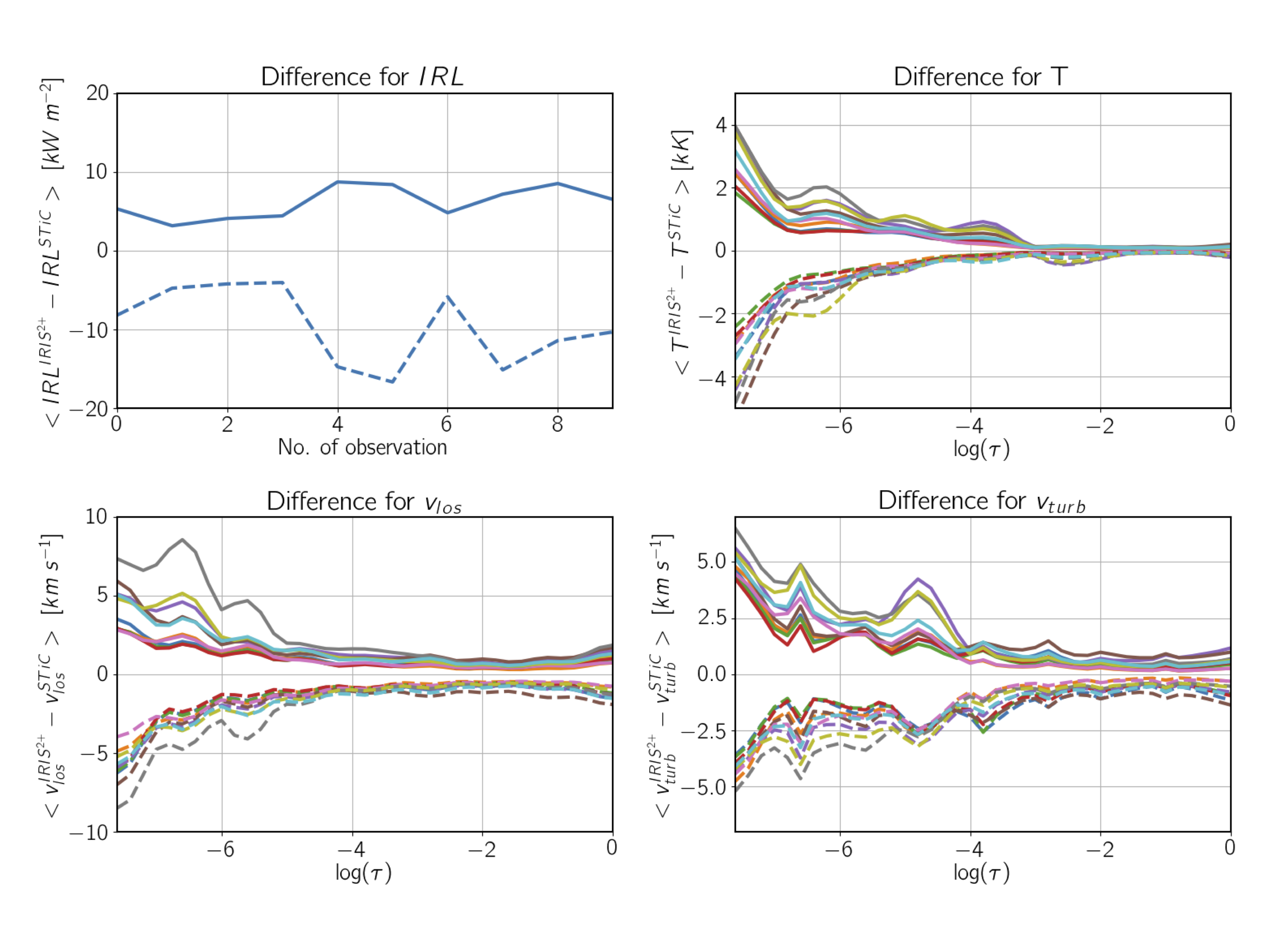}
        \includegraphics[width=0.8\linewidth,trim={10pt 150pt 10pt 150pt},clip]
        {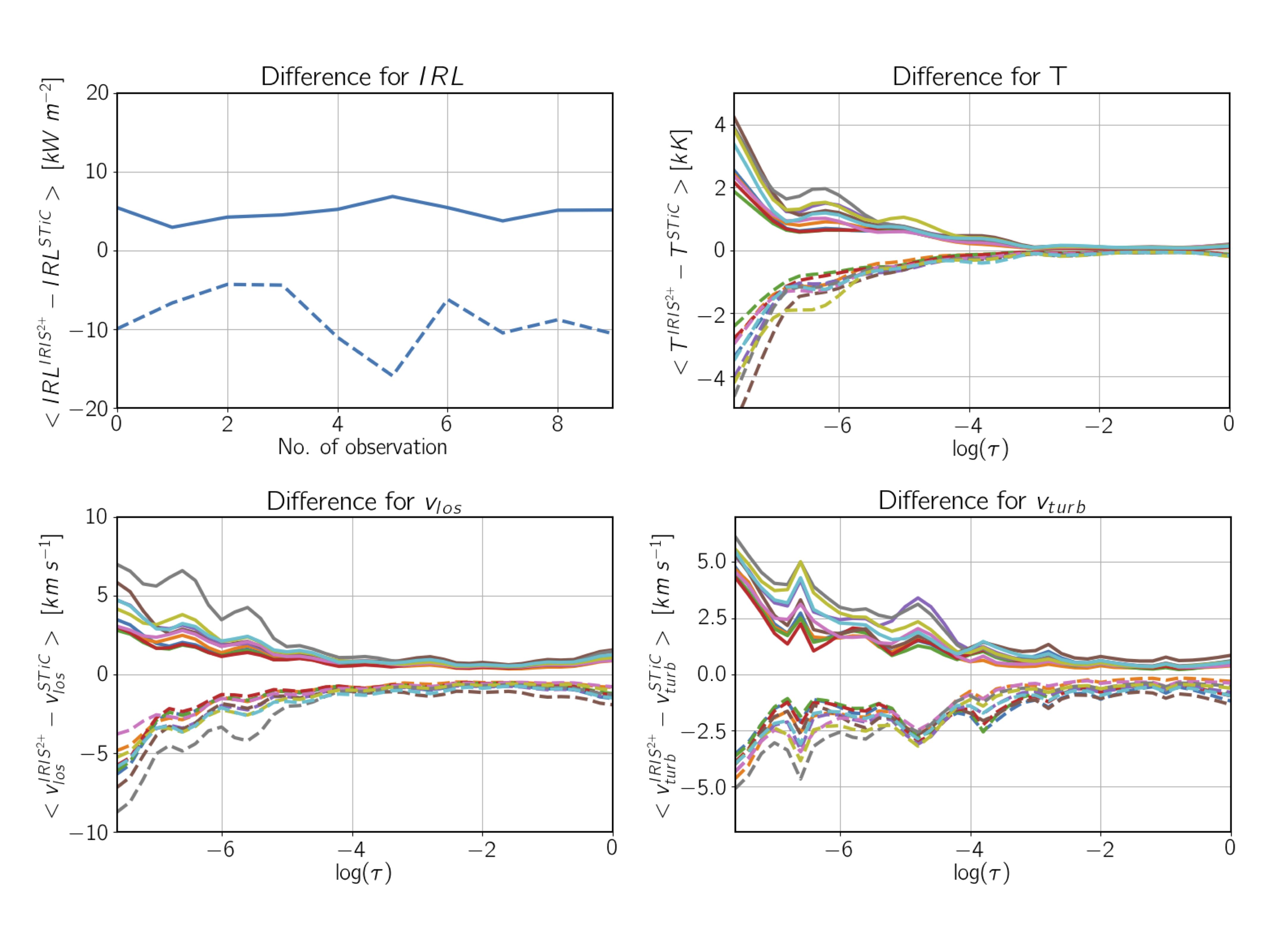}
    \caption{In solid lines, the spatially averaged of the positive differences between the $IRL$, $T$, $v_{los}$, and $v_{turb}$  obtained by \irissqp\ and the same variables obtained by STiC. Similarly, the spatially averaged of the negative differences are shown in dashed lines. The first 4 panels show the differences for the $w_{STiC}$ inversion test, while the last 4 panels show the differences for the $w_{mean}$ inversion test. The color of the lines encodes the number of the datasets used in the study shown in Figure \ref{fig:intmap_1} and Figure \ref{fig:intmap_2}.} 
    \label{fig:plot_absdiff}
\end{figure*}

%



\section{Discussion}\label{sec:discussion}

The IRL obtained by \irissqp\ is comparable to the one obtained by STiC in most of the observations as long as the \cii\ lines and the \mgii\ lines are simultaneously inverted. Considering only the \cii\ lines is not good enough, but considering only the \mgii\ lines may provide acceptable values in comparison with STiC in some cases. Since IRIS always observes the \cii\ lines in some combination with the \mgii\ lines, we can say that \irissqp\ can always provide an IRL as good as the one provided by STiC -- when using the same inversion setup that was used for building the \irissqp\ database.

\asd{The largest difference occurs in the region $-5 < \log \tau < -4$. For some observations, the distribution of T$_{IRIS^{2+}}$ shows an additional component compared to that of T$_{STiC}$, particularly evident at \ltau~=~-4 and \ltau~=~-1, shown in the bottom panels of Figures \ref{fig:lincorr_T_1} and \ref{fig:lincorr_T_2} respectively. As a consequence of this bimodal distribution, the temperature maps at these optical depths obtained with \irissqp\ are hotter than the ones obtained with STiC. 
There are two possible explanations for this behavior. One possibility is that the \irissqp\ database lacks RPs–RMAs capable of reproducing certain features in those observations. As a result, \irissqp\ may fit the observed profiles using RPs associated with RMAs of higher temperature. If this is the case, all RP–RMAs identified by \irissqp\ would need to exhibit temperatures higher than those obtained by STiC within that optical depth range. As mentioned earlier, the overestimation of temperature by \irissqp\ relative to STiC is statistically as probable as underestimation. Therefore, the presence of this extra component must have a different origin.}

The profiles associated with this hotter component are primarily located in the plage regions. A detailed visual inspection of the inversions—including the observed profile, the \irissqp\ fitted profile, the STiC fitted profile, and the corresponding T($\tau$) distributions—reveals an interesting pattern. The quality of the fits for the \cii\ and \mgii\ lines is generally similar between both methods, whereas the fits for the \mguv\ lines are often better in the \irissqp\ inversions than in those from STiC. The opposite tends to occur for the photospheric lines, where STiC typically achieves better fits than \irissqp, but not always. This may explain why the relative number of occurrences in the maximum of the hump in the T$_{IRIS^{2+}}$ distribution with respect to occurrences in the T$_{STiC}$ distribution in the photosphere (see bottom panel of Figure \ref{fig:lincorr_T_2})
is smaller (400:200) than in the mid-low chromosphere (1200:300, see bottom panel of \ref{fig:lincorr_T_1}). This behavior may vary with a different choice of weighting among spectral lines, but it is consistent with findings reported by \cite{SainzDalda19}: the solution obtained by \irissqp\ can sometimes outperform that of STiC. This is because, in this comparison, each observed profile inverted by STiC represents the best result from five randomly initialized inversions. In contrast, \irissq\ and \irissqp\ search through a database of inversion results, each obtained from five independently initialized runs. Consequently, \irissqp\ has a higher likelihood of finding a better overall fit than an STiC single best inversion.

It is worth noting that we are using the STiC inversion as the ``ground truth'', but it would be likely more accurate to consider it as the ``baseline'' we use as a reference to quantify the performance of \irissqp\, with the possibility that \irissqp\ outperforms or underperforms with respect to that reference.

The \vlos\ obtained in the high and mid chromosphere is similar to the values found by STiC, but rather different in the low chromosphere and below (-4$<$\ltau). The frequency distribution of the values obtained by \irissqp\ is narrower than for STiC, that means, \irissqp\ provides slightly lower values of $v_{los}$ than STiC. This might be mitigated by including more atmospheric conditions needed in the \irissqp\ database. 

The \vturb\ recovered from the high to the low chromosphere is similar to the one obtained by STiC, although in general, somewhat lower values are found by \irissqp than by STiC.

\section{Conclusions}\label{sec:conclusions}

\irissqp\ is an inversion tool that provides the thermodynamic conditions from the top of the chromosphere to the mid photosphere, and the radiative losses in the chromosphere, through the inversion of spectral lines that are simultaneously observed by IRIS in the low solar atmosphere. The results obtained by \irissqp\ are much less computationally intensive and in many cases comparable in quality to the inversions obtained by the state-of-the-art STiC inversion code. This code is the one used to build \irissqp\ database, the core of the \irissqp\ inversion tool. 

The main advantages of \irissqp\ over STiC are: the simplicity of use and computational speed. \irissqp\ is totally functional in Python, and it uses standard packages freely available to the public. Although \irissqp\ provides the option to the user to use different weights of the spectral lines that are to be inverted, it also works automatically using well-balanced weights based on the mean values of the integrated intensity of the lines considered for the inversion. In addition to the standard Python packages and a Python interpreter, the only input that \irissqp\ strictly requires is an IRIS raster data file. \irissqp\ will calibrate the input spectra to physical units (for those spectral lines that are present in the \irissqp\ database), transform the synthetic representative profiles (RPs) in the database to the spectral sampling of the input data, crop the lines present both in the input data and the \irissqp\ database to the same wavelength range, train a {\it k-nn} model on this transformed database, predict the closest synthetic RPs to the observed profiles, and finally assign the representative model atmospheres (RMAs) and the integrated radiative loss (IRL) to the location in the raster scan of the observed profiles. The most expensive computational tasks are the calibration of the data and the training of the k-nn model. As in \irissq, the total execution time to invert an IRIS data set is many orders smaller ($O^5-O^6$) than the time required by STiC for the same data set. Typical raster scans can be inverted by \irissqp\ in a matter of minutes.

\irissqp\ also provides an interactive tool to inspect both the observed and the fit profiles, the thermodynamic conditions along the optical depth ($\tau$), i. e.,  T($\tau$), \vlos($\tau$), \vturb($\tau$), and \nne($\tau$), the integrated radiative loss in the chromosphere (IRL), as well as the quality of the fit, given by the $\chi^2$ map.

In summary, when all the \irissqp\ lines (6 chromospheric and 6 photospheric) are inverted simultaneously either by \irissqp\ or by STiC, the results are very similar for the IRL and for the temperature. For the \vlos\ and the \vturb, there are significant differences; usually, lower values are found by \irissqp\ with respect to STiC at some optical depths, mostly in the temperature minimum region and in the photosphere. This indicates that a larger database is needed to increase the atmospheres with different behaviors at some optical depths with respect to others, to decrease the degeneracy of the atmospheres considered in the database.
It is worth noting that, for example, at optical depths corresponding to the top of the chromosphere and the minimum of temperature, the temperature is not well constrained. But this happens both in the STiC results and in the \irissqp\ ones. Therefore, the comparison at these optical depths will be uncertain, as so the inversion results. This behavior is independent of what inversion code was used, and depends on the observed lines and their sensitivity to recover the changes in the thermodynamics along the atmosphere. In other words, we have obtained similar results with STiC and \irissqp\, and the broad dispersion of the results in some optical depths is due to the {\it limited} sensitivity of the lines used in the inversions.
All in all, we have demonstrated that the \irissqp\ results are as good as STiC can provide in most cases.

We believe that increasing the database to 300,000-500,000 RP-RMA-IRL elements would improve the coverage of the conditions present in the low solar atmosphere. To predict the closest RP to an observed RP in a database with such a large number of samples is presently not a challenge for already available tools that are specialized in searching for similarities in large databases of high-dimensional samples, such as the Facebook AI Similarity Search \cite[FAISS, ][]{Johnson19, Douze24}. We have successfully conducted tests with dummy samples similar to the RPs in the \irissqp\ database, proving the feasibility of this suggestion. 

The comparison carried out in this paper uses the results of the $3^{rd}$ inversion cycle of a total of 4 cycles used to build the \irissqp\ database. This is the default option for the \irissqp\ inversion tool. This $3^{rd}$ cycle used 9 nodes for the T, and 8 nodes both for the \vlos\ and the \vturb. Using the database resulting from the $4^{th}$ cycle would provide better fits, but also more variable (complex) thermodynamic values along the optical depth, since it used 13 nodes in T, \vlos, and \vturb.
The option of using \irissqp\ with the results from the $4^{th}$ inversion cycle will also be available at user discretion.

\section{Software and third party data repository citations} \label{sec:cite}

The \irissqp\ inversion tool and \irissqp\ database are available in the IRIS-LMSALpy repository at  \href{https://gitlab.com/LMSAL_HUB/iris_hub/iris_lmsalpy}{https://gitlab.com/LMSAL\_HUB/iris\_hub/iris\_lmsalpy}

The \irissqp\ inspection tool is based on the Python standard 2D graphics library {\tt Matplotlib} \citep{Hunter07}. We have created a Python object ({\tt datacube\_inspector}) that allows the simultaneous inspection of several 3D datacubes that share the size of the first and second dimensions. This is the case for the variables provided by \irissqp.

We have used several machine learning tools contained in the {\tt Scikit-learn} package \citep{Pedregosa11}, and other common packages such as {\tt Numpy} \citep{Harris20} and 
{\tt Scipy} \citep{Virtanen20}.

All these tools use Python \citep{vanRossum95, vanRossum09} as the primary programming language.


\section*{Acknowledgments}
    ASD thanks Souvik Bose and Kyuhyoun Cho for using the preliminary versions of \irissqp\ code: their comments help him to improve it. The authors gratefully acknowledge support from NASA contract NNG09FA40C (IRIS). IRIS is a NASA small explorer mission developed and operated by LMSAL, with mission operations executed at NASA Ames Research Center, and major contributions to downlink communications funded by ESA and the Norwegian Space Agency. JdlCR gratefully acknowledges funding by the European Union through the European Research Council (ERC) under the Horizon Europe program (MAGHEAT, grant agreement 101088184). The Institute for Solar Physics is supported by a grant for research infrastructures of national importance from the Swedish Research Council (registration number 2021-00169). 

%

\vspace{5mm}
\facilities{IRIS}


\software{STiC, \irissqp, Matplotlib, Numpy, Scipy, Scikit-learn, Stats, and other Python libraries were used in this investigation.}



\newpage
 \appendix
 \section{Comparison between the radiative losses calculated including all the species together in non-LTE and calculating them for each atom independently}\label{appendix:comparison_IRL}

We used one of the ARs included in the 	\irissqp\ database (IRIS OBSID \href{https://www.lmsal.com/hek/hcr?cmd=view-event&event-id=ivo\%3A\%2F\%2Fsot.lmsal.com\%2FVOEvent\%23VOEvent_IRIS_20231207_023819_3690108077_2023-12-07T02\%3A38\%3A192023-12-07T02\%3A38\%3A19.xml}{20231207\_023819\_3690108077}) to quantify the effects of calculating radiative losses while considering all species (contributors) as active, i.e., in non-LTE, and when they are considered individually, i.e., with one contributor in non-LTE and the other contributors in LTE. Using STiC, we calculated the integrated radiative losses for the case where all contributors are active ($IRL_{tog}$) for the RMAs of that observation, and when the contributors are considered individually ($IRL_{ind}$), which are the values in the official 	\irissqp\ database. Figures 
\ref{fig:irl_tog_ind_allatmos} to 
\ref{fig:irl_tog_ind_lower} compare these values within the integrated atmosphere defined by optical depths where $4 \leq T~[kK] \leq 18$ at each RMA (labeled as {\it allatmos}), and for the {\it upper, middle}, and {\it lower} partial regions, which are the three equal-sized divisions within that optical depth range.

We selected this dataset because it includes all relevant solar features in the 	\irissqp\ database: umbra, penumbra, plage, and quiet sun. In some RMAs, the convergence when considering all contributors together failed, resulting in zero values. In other cases, the radiative losses were extremely high. These RIRLS were removed from our analysis. From the remaining valid RIRLs, we constructed the corresponding IRL maps, identifying locations associated with the RP-RMA in the data ({\it k-map}). Using these maps, we performed the comparison as described throughout this paper.

The Pearson correlation coefficient for the outlier-filtered data indicates a very high linear correlation in all cases. The slope shows that $IRL_{ind}^{allatmos}$ overestimates $IRL_{tog}^{allatmos}$ by approximately 15\%. In the partial regions of the chromosphere (see \ref{sec:irl} and Figure \ref{fig:irl}), $IRL_{ind}^{upper}$ underestimates $IRL_{tog}^{upper}$ by about 15\%, while for the lower part, $IRL_{ind}^{lower}$ overestimates $IRL_{tog}^{lower}$ by roughly 40\%. This larger discrepancy likely results from limited information from the lower region, closest to the minimum temperature zone, where considering contributors independently misses additional data from other elements. 

Interestingly, for $IRL_{ind}^{mid}$, the Pearson correlation coefficient is the highest after removing outliers, and the slope is nearly 1. This indicates that in the middle part —specifically, the mid chromosphere— the $IRL_{ind}$ estimates the $IRL_{tog}$ very accurately.

In summary, compared to $IRL_{tog}$, the $IRL_{ind}$ provides a good estimate of the IRL in the upper and lower chromosphere, and an excellent estimate in the mid chromosphere. Therefore, the IRL values provided by \irissqp\ must be considered similarly.

 \begin{figure*}
    \centering
    \includegraphics[width=1.0\linewidth]
    {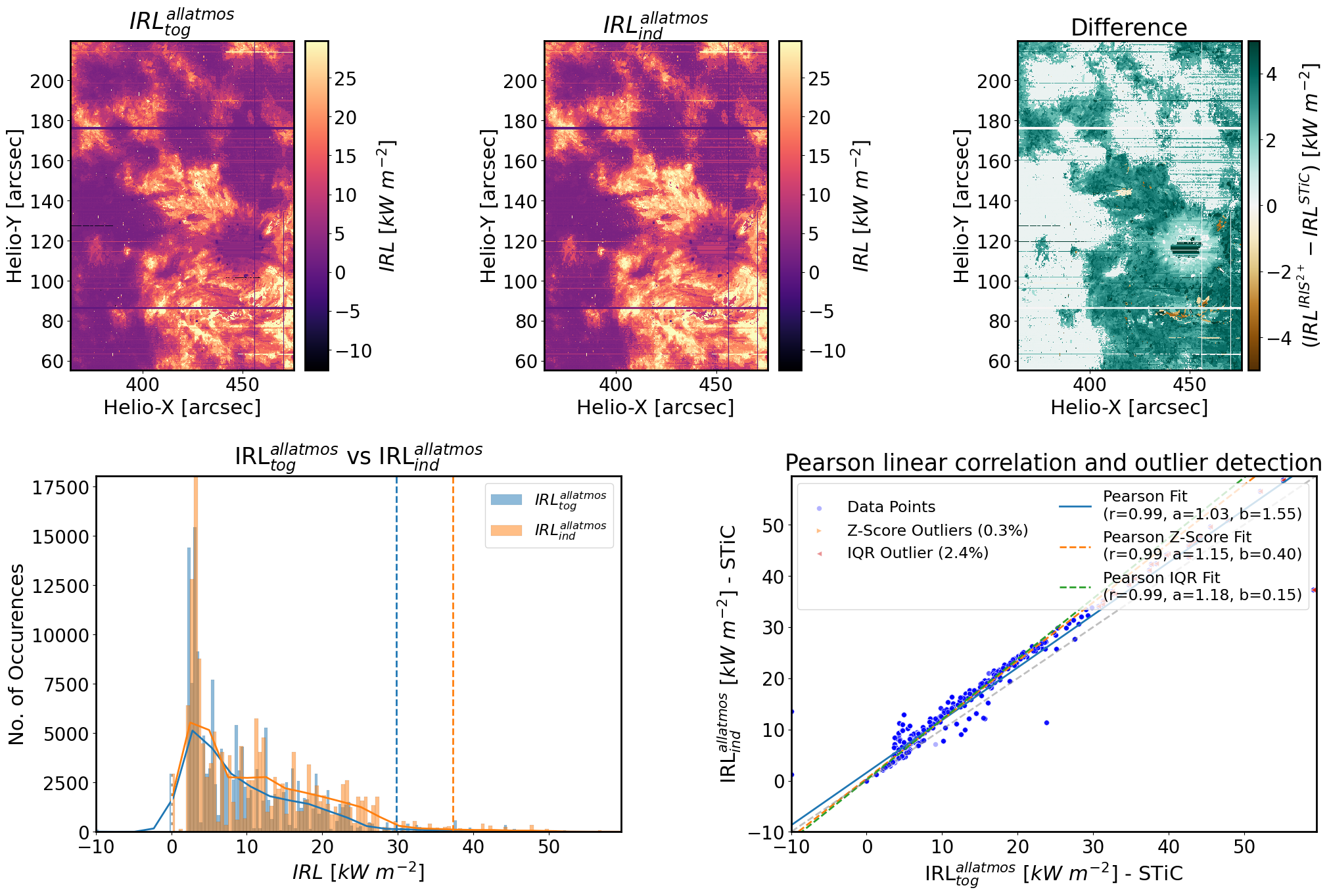}
    \caption{Comparison between the radiative losses computed with all the contributors active (IRL$_{tog}^{allatmos}$) and with the contributors independently active (IRL$_{ind}^{allatmos}$) integrated in the all the chromosphere, i.e., between those optical depths where 18~$<$~T~[kK]~$<$~4.}
    \label{fig:irl_tog_ind_allatmos}
\end{figure*}

\begin{figure*}
    \centering
    \includegraphics[width=.87\linewidth]
    {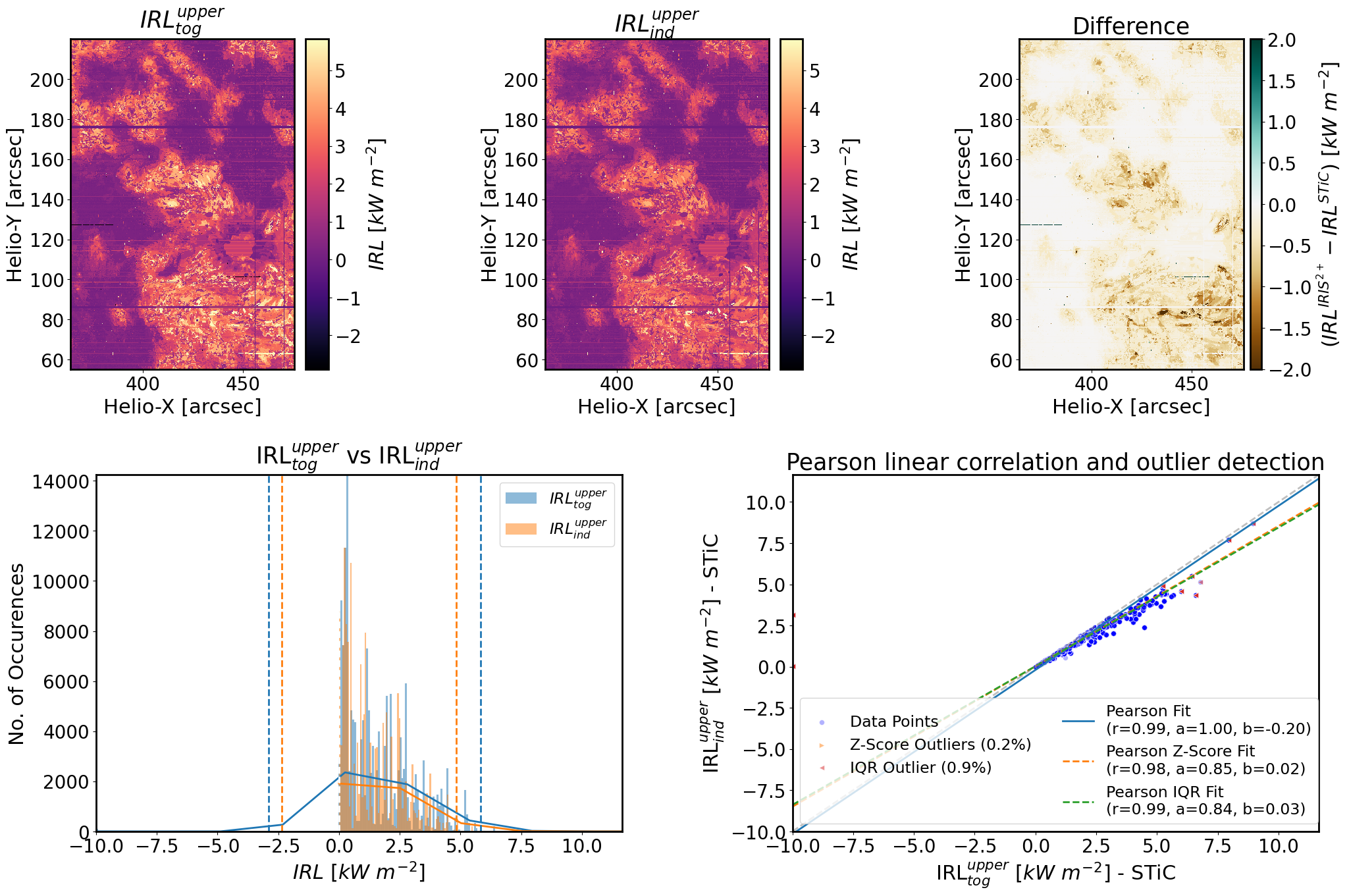}
    \caption{Similar to Figure \ref{fig:irl_tog_ind_allatmos} for the upper chromosphere.}
    \label{fig:irl_tog_ind_upperl}
    
\end{figure*}

\begin{figure*}
    \centering
    \includegraphics[width=.87\linewidth]
    {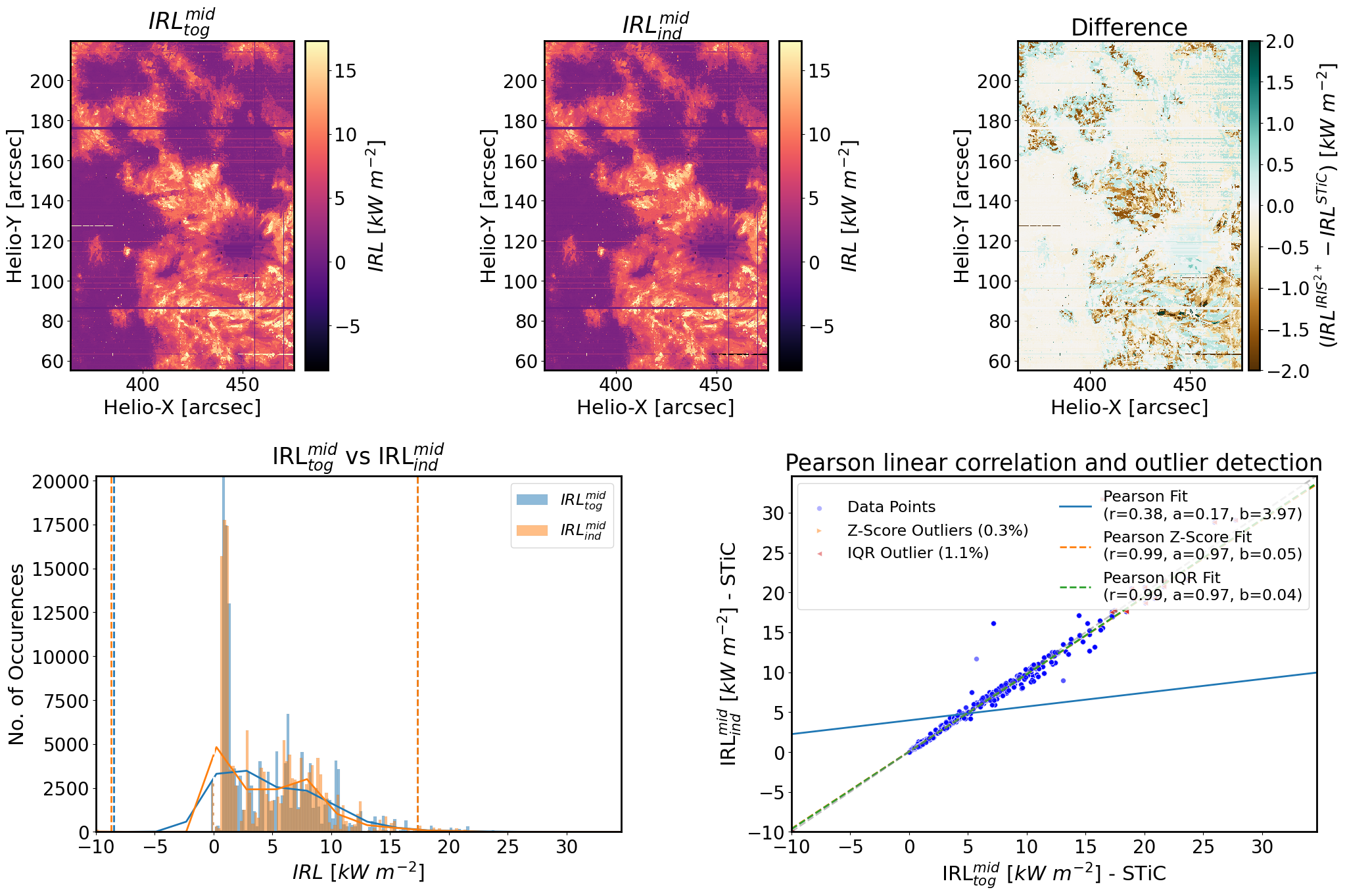}
    \caption{Similar to Figure \ref{fig:irl_tog_ind_allatmos} for the mid chromosphere.}
    \label{fig:irl_tog_ind_mid}
\end{figure*}

\begin{figure*}
    \centering
\includegraphics[width=.87\linewidth]
    {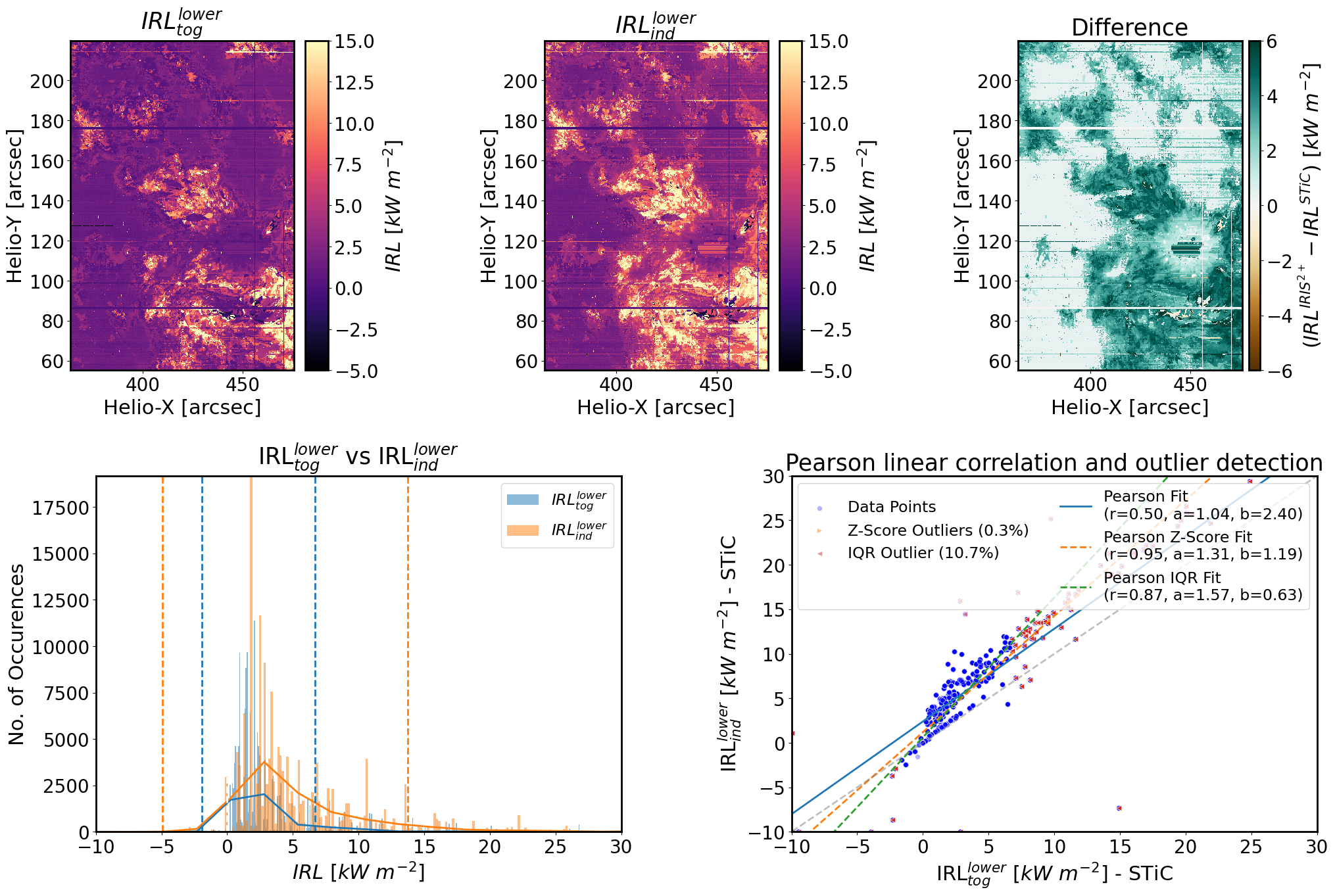}
    \caption{Similar to Figure \ref{fig:irl_tog_ind_allatmos} for the lower chromosphere.}
    \label{fig:irl_tog_ind_lower}
\end{figure*}

 \section{Comparison of $\chi^2$ and the role of the weights}\label{appendix:detailed_comparison_chi2}
 
 To study the impact of different selections of the \irissqp\ lines during the inversion, we have made two main comparison experiments: i) several combinations of weights are used when all the lines are simultaneously inverted, and ii) a selection of lines is inverted with \irissqp\ using the same weights as those used with the STiC inversion. In the latter case, it is important to note that this comparison is between the results obtained by \irissqp\ inverting only the selected lines with the results of STiC inversions, which always consider all the lines. Thus, in the second experiment, we compare inversions based on {\it only a few lines} with the best inversions obtained by inverting all the lines simultaneously with STiC. 
 
 We first compare the $\chi^2$ obtained from the STiC and IRIS, then the radiative loss, and finally the thermodynamic values. We use the results of the \irissqp\ inversion considering all the lines using the same weights used by STiC of the data set \#0 to illustrate the comparison procedure we have followed with all the other data sets and all the experiments run in this investigation.
  
Our inversion problem consists of evaluating goodness of fit in a very high-dimensional domain, where the observed profiles and the synthetic RPs in \irissqp\ database sample 944 spectral wavelengths.

A critical aspect of inversions, and therefore of the $\chi^2$ goodness-of-fit criterion, is to determine the weights for the different spectral features in the sampled data. In our case, this means the respective weights applied to the spectral lines that are included. The idea behind using weights is to scale the lines in such a way that their contribution to the $\chi^2$ is balanced, or increased or decreased with respect to other lines. Thus, if we want to discard a line(s) in the fit between the observed line and the synthetic profile during the inversion process, we will impose a weight that is equal to zero for the wavelengths corresponding to that line(s). On the other hand, if we want to emphasize the importance of a particular line(s) or, in some cases, some part(s) of the line(s) (e.g., the core {\it versus} the wings), we can assign a higher weight to these wavelengths, i.e., a higher value to the core or a smaller value to the wings in our example. 

For computational reasons, only one array of weights at each wavelength is given for all the observed profiles included in a data set. 
This is not a problem if the relative variation of the intensity of the lines along the data set is negligible, but it can become an issue if that relative variation is important. In that case, it is advised to sort the data into different batches, each of one having its own combination of weights. This is the strategy that \citep{SainzDalda23}
followed to simultaneously invert the \cii, \mguv, and \mgii\ lines in the flare ribbons during the maximum of an X-class flare, where the ratio between the \cii\ lines and the \mguv\ and the \mgii\ lines in the ribbons was very different from ratios found outside the ribbons. Such a dedicated strategy has not been applied to \irissqp, since flares are not included in our database yet. 

As we already mentioned in Section \ref{sec:inversion}, while building the \irissqp\ we found a combination of weights that is valid for most of the profiles in the database. 

We have studied the impact of the selection of weights for the different lines when all the lines in the \irissqp\ are simultaneously inverted. To this aim, we have considered the following combinations of spectral lines and weights:

\begin{itemize}
    \item {\it All lines $w_{STiC}$}: the weights passed to the \irissqp\ inversions are strictly the same as the
    ones used for the STiC inversions. All the spectral lines that are available within \irissqp\ are fit. 
    \item {\it All lines $w_{mean}$}: the weights for the \irissqp\  inversions are set as the mean value of the integrated intensity for each line with respect to the mean value at the \mgiik\ line. This is the ``{\tt mean}'' weighting option mentioned above.
    \item {\it All lines $w_{default}$}: the relative weights are set as follow: $w_{\textrm{\cii}}$:$w_{\textrm{\mguv}}$:$w_{\textrm{\mgii}}$:$w_{\textrm{photos}}$ = 100:3:1:2. This is the already mentioned ``{\tt default}'' weighting option of \irissqp. 
\end{itemize}

For the case that either the user selects a few lines of the available ones in a dataset, or only some lines of the \irissqp\ database are observed in a particular dataset, we have considered a combination of lines attending to the possible interest of the user in a particular region of the solar atmosphere. In the following cases, for a better comparison with the results obtained with STiC, the weights given to the lines in the \irissqp\ inversions are the ones used in the STiC inversions. These are the studied combinations:

\begin{itemize}
    \item {\it \cii + Mg UV + \mgii}: simultaneously consider the \cii\ lines, all the lines of the \mguv, and the \mgii. These lines are sensitive to changes in the thermodynamics from the top to the bottom of the chromosphere. 
    \item {\it \mgii}: only the \mgii\ lines, including the \mguvtt\
    lines located between them are considered. These lines are sensitive to changes in the thermodynamics from the mid to the bottom of the chromosphere.
    \item {\it \cii}: only the \cii\ lines are considered. These lines are sensitive to changes in the thermodynamics at the top of the chromosphere.
    \item {\it Photospheric}: only the photospheric lines are fit. These lines are: Ti \iifns\ 2785.46 \AA, Fe \ifns\ 2793.22 \AA, Fe \ifns\ 2809.15 \AA, C \ifns\ 2810.58, Ni \iifns\ 2815.18 \AA, and Fe \ifns\ 2827.33. These lines are sensitive to changes in the thermodynamics from the bottom of the chromosphere to the mid-photosphere.
\end{itemize}

The goal of these experiments is to quantify how well inversions based on a subset of lines or a different choice of weights can recover the information provided by inversions that include all lines simultaneously. 

\begin{figure*}
    \centering
    \includegraphics[width=1\linewidth]
    {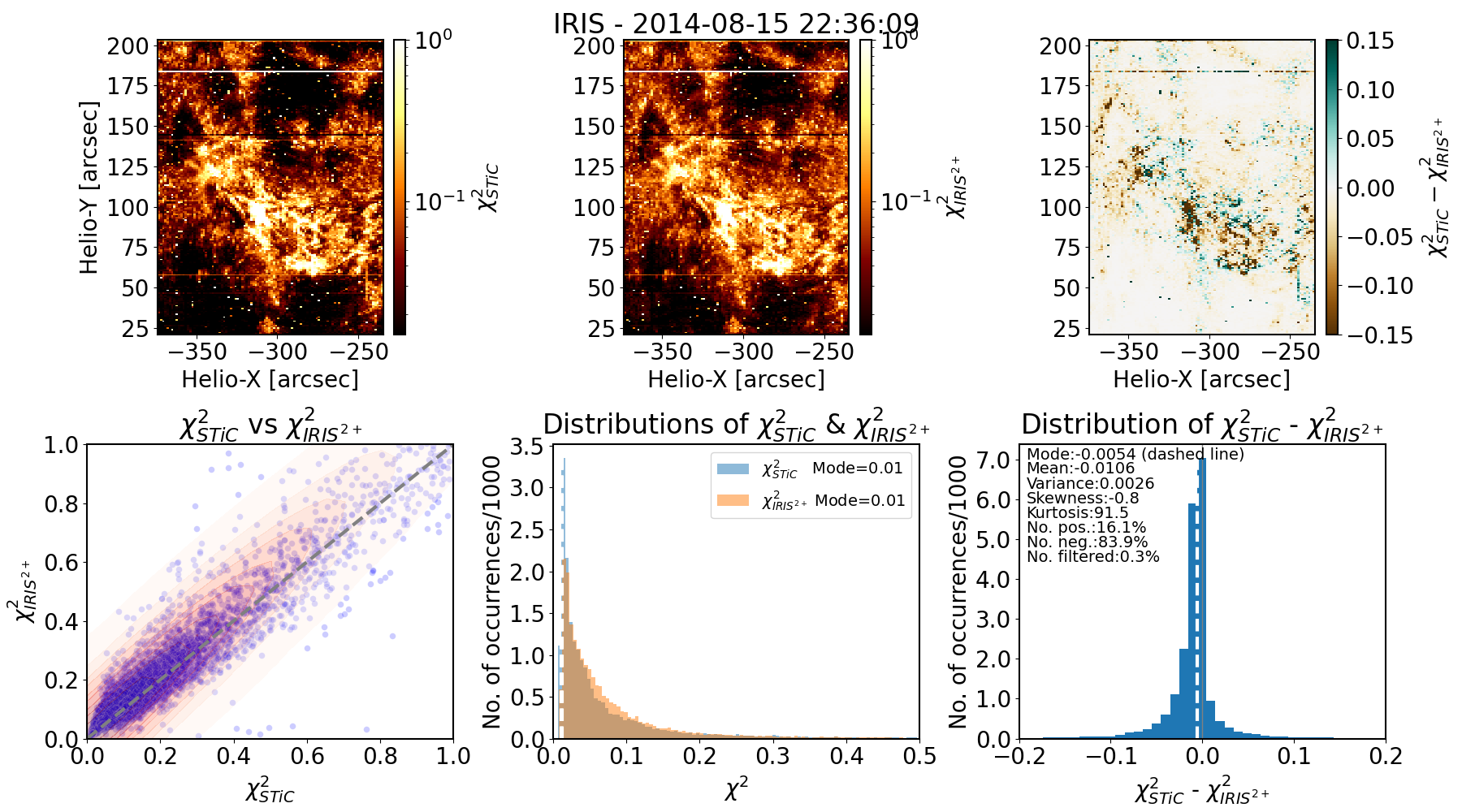}
    \caption{Comparison between \chistic\ and \chiirissqp\ for data set \#0 shown in Figure \ref{fig:intmap_1}. The top row shows the $\chi^2$ maps for the inversion considering all the spectral lines and the weights that are used in the STiC inversion, both for an inversion with STiC ($\chi^2_{STiC}$, left) and \irissqp\ ($\chi^2_{IRIS^{2+}}$, middle), i.e., the case we call ``{\it All lines $w_{STiC}$}''. The right panel in the top row shows the difference between $\chi^2_{STiC}$ and $\chi^2_{IRIS^{2+}}$.  The plots in the bottom row show the frequency distribution (left) and the scatter plot (center) for 
    $\chi^2_{STiC}$ and $\chi^2_{IRIS^{2+}}$, and the frequency distribution for their difference (right).}
    \label{fig:mapchi2_1_2}
\end{figure*}
\begin{figure*}
    \centering
    \includegraphics[width=1\linewidth]
    {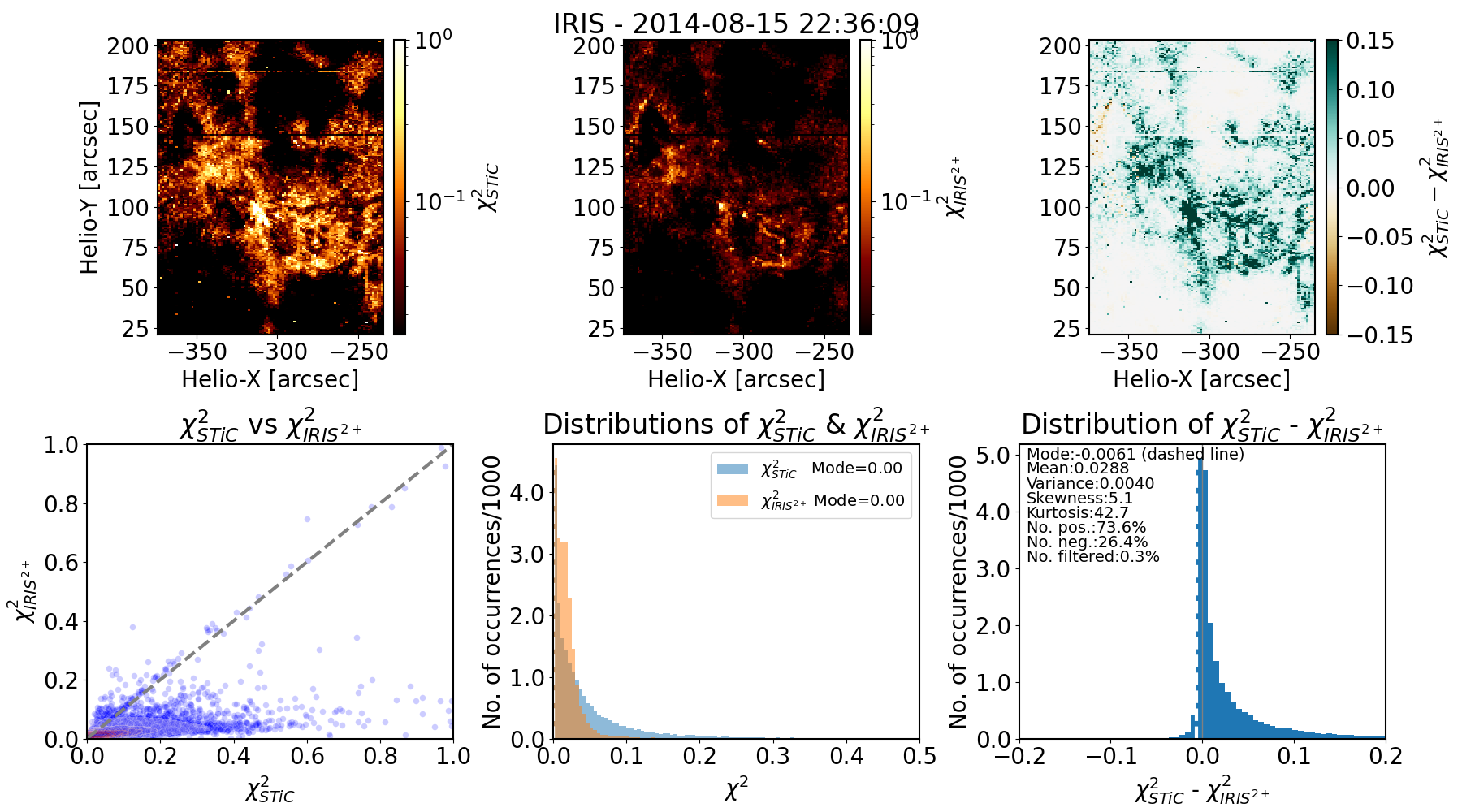}
    \caption{Same as Figure \ref{fig:mapchi2_1_2} for the inversions considering only the \mgii\ lines (``{\it \mgii\ \& UV23}'' case). See text for details.}
    \label{fig:mapchi2_2_2}
\end{figure*}

Figure \ref{fig:mapchi2_1_2} shows a comparison between \chistic\ and \chiirissqp\ for observation \#0 of the selected data sets, and the difference between them (left, middle, and right panels in the top row, respectively). 
In the scatter plot (left panel, bottom row), all the values of \chistic\ (on the x-axis) are compared to the values of \chiirissqp\ (on the y-axis), showing a clear linear relationship. 
In this particular observation, the distribution of the $\chi^2$ values (middle panel in the bottom row) is 
very similar, although the \chiirissqp\ values are slightly larger than those found for \chistic\. Both distributions have very similar mean and mode (i.e., the value that occurs most often). This fact can also be seen in the distribution of the difference of the $\chi^2$ values (right panel).
This means, in general, for the inversion of this observation, that \chiirissqp\ $>$ \chistic. 

The comparison shown in Figure \ref{fig:mapchi2_1_2} refers to the inversion case of \irissqp\ using the same weights that were used for the STiC inversion, i.e., the {``\it all lines $w_{STiC}$''} case. Figure \ref{fig:mapchi2_2_2} shows the comparison for the case {``\it \mgii''} which means that \irissqp\ only fits the \mgii\ lines, but uses the same weights as those used by STiC for these lines. As we can see, now \chistic$>$\chiirissqp\ in most of the pixels in the data. Note that although the calculation of \chistic\ only takes into account the \mgii\ lines, the STiC inversion fits all the lines. Nevertheless, it is not surprising that \irissqp\, only focused on the \mgii\ lines and ignoring all the other lines, provides a better fit for the former lines. As a consequence, the thermodynamic state recovered in the mid and low chromosphere by \irissqp\ may be more accurate than that provided by STiC, which considers all the lines. We note however, that for this case of \irissqp\ the thermodynamics are not (as) well constrained outside of the mid and low chromosphere. 

\begin{figure}
    \centering
    \includegraphics[width=1\linewidth]
    {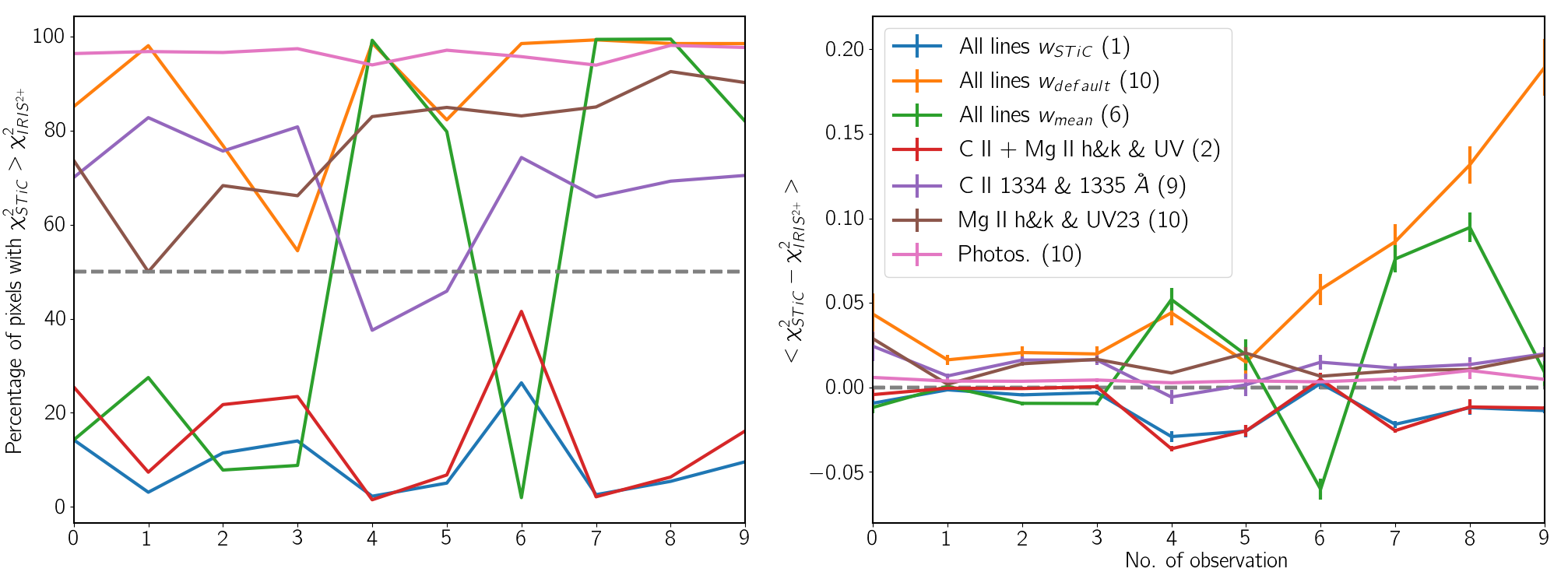}
    \caption{Left: percentage of pixels with \chistic\ $>$ \chiirissqp\ for all the datasets and the inversions carried out with different selections of lines and weights. Right: average difference between the \chistic\ and \chiirissqp\ for the same inversion experiments.}
    \label{fig:stats_chis2}
\end{figure}

\begin{figure*}
    \centering
    \includegraphics[width=1\linewidth]
    {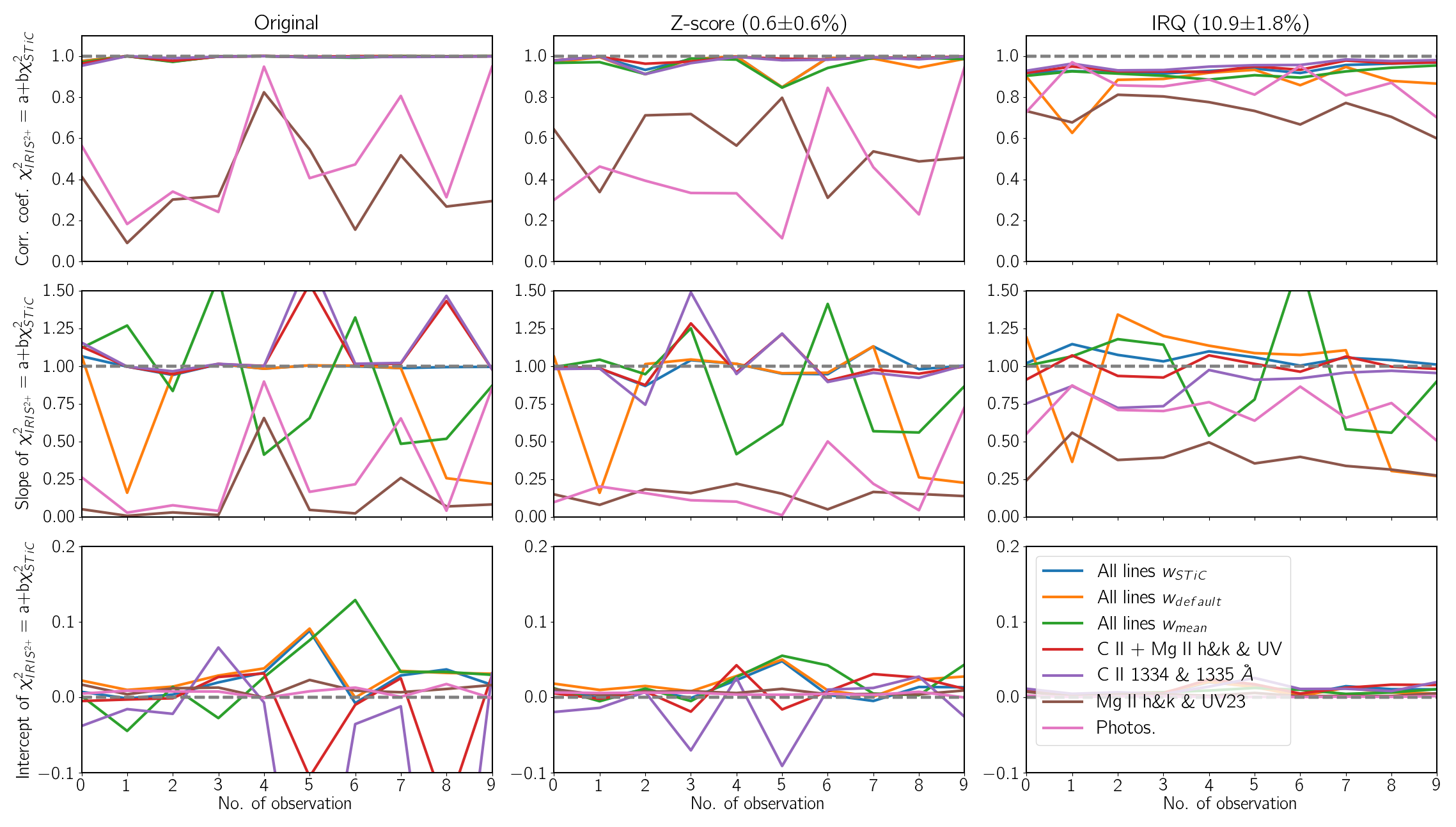}
    \caption{Linear correlation parameters between \chistic\ and \chiirissqp for all the data sets for all the weighting experiments.}
    \label{fig:lincorr_chi2}
\end{figure*}

We have made similar plots for all the selected data sets and for all the combinations of weights detailed above to visually inspect the behavior of \chistic\ and \chiirissqp. These plots provide us information about the \chistic\ and \chiirissqp\ distributions and the linear relationship between these variables. However, the overall results of this comparison are better presented in Figures \ref{fig:stats_chis2} and \ref{fig:lincorr_chi2}. Figure~\ref{fig:stats_chis2} shows for each selected data set the percentage of pixels with \chistic $>$ \chiirissqp\ (top panel), and the mean of the difference between \chistic and \chiirissqp\ (bottom panel). As we can see, the number of pixels with 
\chistic $>$ \chiirissqp\ is less than 50\% for all the datasets inverted by \irissqp\ using the same weights as used for the STiC inversions, both for all the lines (blue line) and for the chromospheric lines (red line), and for half of the data sets inverting all the lines with \irissqp\ using the ``{\tt mean}'' weighting method (green line). 

When \irissqp\ uses the ``{\tt default}'' weighting method,  \chistic $>$ \chiirissqp\ for more than 50\% of all the datasets. This is because the relative imposed weights for the \cii\ lines in the \irissqp\ inversions using the ``{\tt default}'' weighting are in all cases smaller than the ones used in the STiC inversions. Therefore, the
fitting of these lines during the \irissqp\ inversions is less ``important'' than it is in the STiC inversions. As a consequence, the best fit found by \irissqp\ is, in most cases, worse than the one found by STiC.

For the ``{\it \cii}'' (purple), ``{\it \mgii}'' (brown), and ``{\it Photospheric}'' (magenta) experiments, which use the same weights in \irissqp\ as in the STiC inversions, but only for the selected lines, all the datasets have more than 50\% of the pixels with \chistic $>$ \chiirissqp. 

 The bottom panel of Figure \ref{fig:stats_chis2} shows the average of the difference between \chiirissqp and \chistic. As we can see, for many datasets, the difference is very small. The numbers in parentheses next to the labels in the legend indicate how many datasets have, on average, a difference larger than 0, which means that on average the \chistic $>$ \chiirissqp\. The case for which \irissqp\ is using the mean value for the relative weights for all the lines (green line) shows an interesting behavior. For 6 data sets, the \chistic $>$ \chiirissqp\. This may occur because either the weights used by \irissqp\ are rather different from the ones considered by STiC or because they are similar and the fit is better for the \irissqp\ inversions. Interestingly, in the first 4 data sets, the difference is very small. We can conclude that in some cases \irissqp\ inversion fits are almost as good as the ones obtained by STiC, while in the others \irissqp\ may provide a better solution, or in a few cases significantly worse, e.g., for the data set \#6.

Figure \ref{fig:lincorr_chi2} shows the parameters of the linear correlation fit for \chiirissqp = $a+b$\chistic, with $a$ the intercept and $b$ the slope.

The first row of Figure \ref{fig:lincorr_chi2} shows the correlation coefficient. The inversions considering only the \mgii\ lines (brown) and only the photospheric lines (magenta) show a low correlation coefficient. All the other cases show a moderate or large correlation coefficient. However, for these cases, it increases to a high linear correlation once the values have been filtered with the IQR method. With this outlier detection method, all the cases show a large linear correlation between \chiirissqp\ and \chistic. The slope of this relationship is shown in the second row. For the cases inverting all the spectral lines simultaneously (blue, orange, and green lines), the slope is close to 1 in most of the data sets, as well as for the case of the inversion of only the chromospheric lines (red). For the inversion of only the \cii\ lines, only the \mgii\ lines, or only the photospheric lines, the slope is lower than 1. This means, for a given value of \chistic the predicted \chiirissqp\ is definitely lower. This behavior can be seen in Figure \ref{fig:mapchi2_2_2}, both in the $\chi^2$ maps and in the scatter and distribution panels (bottom row). 

\asd{It is worth noting that the \irissqp\ user should pay special attention to those cases where an \irissqp\ spectral 
line is barely above or below the noise level of the data. In these cases, discarding that line, i.e., weighting 
it with 0.0, may return a better fit in the other lines. Another case may be the interest of the user, for instance, to 
recover the thermodynamics in the chromosphere. In this case, weighting the photospheric lines to 0.0 most likely provides a better fit in the \irissqp\ chromospheric lines, and therefore a more reliable chromospheric thermodynamics. In general, the user of \irissqp\ must keep an active and critical attitude towards the weights used or selected during the inversion. \irissqp\ offers several weighting options to favor this habit.}



\bibliography{allbib, others, jaime}{}
\bibliographystyle{aasjournal}

\end{document}
